\documentclass[a4paper,10pt]{article}
\usepackage{ucs}
\usepackage[utf8]{inputenc}
\usepackage{amsmath}
\usepackage{amsfonts}
\usepackage{amssymb}
\usepackage{amsthm}
\usepackage{rotating}
\usepackage{xcolor}
\usepackage[english]{babel}
\usepackage[T1]{fontenc}
\usepackage{graphicx}
\usepackage{hyperref}
\usepackage{cuted}
\usepackage{verbatim}
\usepackage{subfigure}
\usepackage{multirow}
\usepackage{bbold}
\usepackage{slashed}
\usepackage{indentfirst}
\usepackage{tcolorbox}
\usepackage{placeins}
\usepackage[a4paper, total={7in, 10in}]{geometry}

\usepackage{tikz}
\usepackage[compat=1.0.0]{tikz-feynman}

\newcounter{eq}

\title{\bf  U(5) Nambu-Jona-Lasinio model 
 with flavor dependent coupling constants:
pseudoscalar and scalar mesons masses
}
 \author{ W. F. de Sousa$^{1,2}$ , F. L. Braghin$^1$
\\
{\normalsize $^1$ Instituto de F\'\i sica, Federal University of Goias, 
Av. Esperan\c ca, s/n,
 74690-900, Goi\^ania, GO, Brazil}
\\
{\normalsize $^2$ 
Instituto Federal de Goi\'as,
R. 75,  n.46, 74055-110, Goi\^ania, GO, Brazil}
}

\date{}

\begin{document}

\maketitle

\begin{abstract}
By considering the  background field method
we calculate  one-loop polarization corrections to 
the coupling constant of the  flavor-U(5) 
Nambu-Jona-Lasinio (NJL) model with degenerate up and down quarks.
They break flavor and chiral symmetries and they can be written as
$G_{ij}^\Gamma (\bar{\psi} \lambda_i \Gamma \psi) 
( \bar{\psi} \lambda_j \Gamma \psi)$,
for the scalar and pseudoscalar channels
($\Gamma=I , i \gamma_5$) and $i,j = 0,1,...,N_f^2-1$.
Therefore, these coupling constants
 do not introduce further free parameters in the model which amount to five, six or seven in total.
Their contributions to different observables are computed:
 quark-antiquark scalar condensates, 
masses of  quark-antiquark meson states (pseudoscalar  and scalar) and
pseudoscalar meson weak decay constants.
 The non-covariant three dimensional regularization scheme is employed 
leading to an unique interpretation for light and heavy quarks cutoffs.
 Besides that, 
 flavor dependence of cutoffs is implemented in an unambiguous way.
It turns out that only two UV cutoffs ($\Lambda_f$)
are best suitable, one for the 
light ($f\equiv l=u,s$) and one for the heavy quarks ($f\equiv h=c,b$).
Nevertheless,  the best results  are  obtained for nearly
flavor-independent cutoffs. 
A quantum mixing due to the
representations of the flavor group 
leads to different 
interactions for the quarks and for 
the meson states, respectively denoted by $G_{ff}$ and $G_{ij}$.
This  quantum mixing effect is responsible for a lowering of 
quark effective masses and a slight improvement of predictions of 
observables.
 A quite  surprisingly good description of  almost 
all the 25-plet pseudoscalar meson masses
 (within nearly  $5\%$)
and most of the
 scalar meson masses -is obtained 
 within nearly 10$\%$.
However, the usual problems to describe the correct mass hierarchy  of some light 
scalar mesons still remain.  The NJL-gap equations
seems to overestimate 
the heavy quark condensates at the usual mean field level
usually adopted for model.
In spite of the   good description of the meson masses, 
the pseudoscalar meson weak decay constant cannot be described by 
the NJL model
with relativistic heavy quark propagators
without further
 interactions or effects.
\end{abstract}

\section{Introduction}

Whereas light hadron  dynamics respect approximate chiral symmetry due to 
relative smallness of u,d and s quark masses,
 heavy hadrons cannot be argued to  have  chiral symmetry as a 
guiding principle
 for  their dynamics.
However, for the description of  heavy-light mesons, 
there may emerge a difficulty 
 to disentangle the relative contribution of chiral symmetry
and  dynamical chiral symmetry breaking (DChSB), that are important 
for the  light quark sector, from particular properties from heavy quarks
since there are possible relevant symmetries that emerge in the
heavy  quark (HQ)  limit, for  $m_Q \to \infty$,
  for example 
\cite{yan-etal,nowak-etal,HQ-review,review-HQS,isgur-wise,HQ-observ}.
Heavy-light meson spectra were found to present degeneracies
associated to  HQ symmetries, such as in the pseudoscalar and vector 
 or scalar and axial multiplets
\cite{HQS-wise,cheng-etal-S+A}.
 Some   ways of  describing reasonably well
  heavy hadrons  
are 
given by,  for example,   Heavy Quark Effective Field Theory (HQEFT)
and 
 models with Heavy quark symmetries 
including those for  heavy-light mesons
\cite{review-HQS,HQS-wise,nowak-etal,beneke,HQS-ChDyn,review-CL,jiang-etal-2019,yan-etal-1992}.
In spite of the continuous efforts  to obtain results from first principles QCD,
  lattice QCD  \cite{lattice}, 
and continuous approaches in the Euclidean space 
\cite{SDE}, it still is interesting to consider effective models
in which the physical meaning of parameter are usually clear,
also  new effects can be 
proposed and tested.
For more  reliable models, results can lead to the identification of the 
most important degrees of freedom  with grounds in QCD.
  Notwithstanding,
it becomes important to disantangle 
 mathematical features of the model and issues that must be directly 
 traced back to the QCD degrees of freedom
 \cite{strings,PRD-2022b}.

Among the most successful hadron effective models, in particular  for
the low energies regime,
the   Nambu-Jona Lasinio (NJL)   model  and its extensions
\cite{NJL,klevansky,vogl-weise,hatsuda-etal,baryon-NJL1,baryons-2}
are expected to be appropriate whenever  chiral symmetry and its 
DChSB are  relevant.
Standard treatments of the NJL for  flavor U(5), sometimes with modifications  usually 
incorporating
 heavy quark symmetry,
have  also been already considered 
\cite{HM-NJL-guo-etal,npb-ebert-etal-1995,gottfried-etal,nam-prd,bardeen-hill,arriola}.
The model has also been  source of inspiration 
for high energies interactions \cite{tt-btquarks}.
The chiral scalar quark-antiquark condensates 
represent an order parameter for the DChSB and
earlier estimations from QCD sum rules  
suggested  heavy quarks 
scalar condensates  (mostly b and t) may be suppressed
due to the large quark current masses
 \cite{QCDSR-nocond}.
However, heavy quark-antiquark scalar condensates  were found to
not be entirely suppressed 
 \cite{NJL-t-b,anton+ribeiro}.
Therefore  DChSB might  be expected to
provide some contribution for heavy quarks dynamics
in spite of the corresponding large explicit chiral and flavor symmetry breakings in QCD
Lagrangian,  and 
in the NJL model.
One advantage of the  
 model is the possibility  to incorporate (somehow uniformly)
light and heavy hadrons (usually mesons) consistently within
the framework of the quark model.
Indeed, the NJL model can be associated to a 
relativistic version of the   constituent quark model 
that was considered early to investigate heayy hadrons
\cite{godfrey-isgur,relativ-effects}.
In such   appealing framework to describe hadrons in
the corresponding flavor-multiplets, 
that shows direct relation to 
QCD degrees of freedom, it is
 interesting to investigate further its suitability to describe mesons in the 
U(5) flavor  NJL model  as quark-antiquark states
 (with charm and beauty, c and b)
by analyzing possible theoretical or phenomenological  modifications.

The NJL model is non-renormalizable and  an ultraviolet (UV) cutoff
is required to provide predictions for observables.
Several regularization methods have been
employed and compared mostly for the light quark sector and, in this case,
results are nearly equivalent \cite{klevansky,kohyama-etal}.
For the heavy quark sector, it is natural to expect  that
covariant and non-covariant regularization schemes 
should lead to quite different results.
It  is easily noted that  higher energies (larger than the UV cutoff from the NJL model) 
are needed to investigate such heavy quarks/hadrons.
and it  is usually  argued that light and heavy  (l and h) quarks, due to the large mass differences,
are subject to different energy scales and,   in a description of 
punctual  interactions of an effective model,
  flavor-dependent cutoffs   should be considered
\cite{SBK,mexicanos-2019,bashir-etal,craig-etal}.
\footnote{
It  is  not clear how and to what extent these different channels, 
scalar-pseudoscalar channels for the standard NJL and vector  or vector-axial for 
other models,
are fully or partially independent  because of  the
 Fierz transformations (FT)
and of the intrinsic ambiguity in performing  the FT
\cite{fierz-transformation}. 
This issue will not be discussed further in the present work.
}
 Energy scale, or size scale,  parameters might also be important 
in defining  non relativistic and relativistic heavy quark potentials  
 \cite{heavy-quark-pot,relat-heavy-pot}.
However, by employing a  non-covariant (three-dimensional) regularization method the meaning of the 
UV cutoff - acting in the three-dimensional momenta of all quarks - 
might become more uniform,   and maybe natural,  for both heavy and light quarks.
It will be seen
that 
any difference among the light and heavy quark  cutoffs  can be considerably small
in the non-covariant three-dimension regularization scheme.

It has been argued that different quarks, with  different 
constituent masses,
should interact by flavor-dependent coupling constants
for both light and heavy mesons
\cite{review-CL,pattern-interaction,PRD-2021}.
Flavor dependent contact interaction were also envisaged 
for high energy quarks at the LHC energies 
from a power counting approach \cite{contact-lhc}.
Given that the quark current masses are the only flavor dependent parameters in 
QCD action, the parameters of an effective model, 
such as
 punctual effective coupling constants,
should also  incorporate such a  flavor dependence
- similarly to an effective field theory  \cite{kaplan}.
Furthermore, it is  interesting to understand the effects  
of small and large  quark masses in the parameters of the model 
because of  quantum effects of the model itself. 
In the light flavor U(3) NJL model, flavor dependent coupling constants
due to vacuum polarization were 
calculated and tested  in \cite{PRD-2021,JPG-2022}.
In the present work, the
quark-antiquark vacuum polarization mechanism that generate
flavor-dependence of NJL-coupling constants will be  addressed for 
the flavor U(5) NJL model.
It is important to emphasize that these flavor dependent coupling constants
do not introduce further parameters in the model
since they are calculated with the quark effective masses $M_f^*$ and
with  the original coupling constant of 
reference $G_0$.

The large number of hadrons found
in experiments and the difficulties in obtaining experimental information about their properties
make  it difficult to establish the validity of the quark-antiquark picture for 
many mesons.
Which of the heavy mesons observed in experiments belong 
a  quark-antiquark  multiplet  from the quark model
still is somewhat under scrutiny for several channels.
For example, earlier theoretical  estimates for $c\bar{b}, b\bar{c}$ 
pseudoscalar mesons $B_c$ were
done in \cite{bc-1-2} and, currently,
 the pseudoscalar quark-antiquark  meson multiplet - with c and b quarks - 
 is basically  considered settled.
Other non-standard states have been seemingly found 
- for example in 
\cite{exotic,B-mixing,exotic-pseudosca,mass-shifts-channel}.
Besides that, longstanding difficulties are found for the description of several
 (light) scalar mesons  for  which there are indications of different 
components besides the 
 quark-antiquark structure 
\cite{scalars-light,NJL-scalars}.
For the scalar heavy mesons it is not yet possible to identify clearly 
such type of problems.
However,  some investigations indicate some quark-antiquark structures
such as, for example,  
for the heavy-light  
 $D_0^*$
identified  as  $c\bar{q}$ states
when including the contribution of mesons loops    \cite{guo-etal-2008}.
In the multiplet (nonet) of light pseudoscalars 
there was a problem for  describing the states built up with 
 the 
diagonal generators, leading to 
the well known
 $\eta-\eta'$ puzzle.
A  solution for that was found by 
considering the axial anomaly that drives a 
meson mixing \cite{PS-mixing}.
In the flavor-SU(3) NJL model this is usually implemented by a 
sixth  order ( 't Hooft)  determinantal interaction
 \cite{thooft,bernard-etal-1987,dimitrasinovic-hl},
although flavor dependent coupling constants induced by 
vacuum polarization
also induces the corresponding mesons mixing \cite{PRD-2021,JPG-2022}.
The
determinantal 't Hooft interaction, that would be a tenth order 
interaction for flavor U(5)  \cite{creutz}, 
 will not be considered in the present work so 
that the effects of the flavor-dependent coupling constants
induced by vacuum polarization can be isolated and understood separately.
In the present work, we will adopt a quark-antiquark structure for the scalar mesons 
that is compatible with the standard quark model and usually discussed in the literature.
 In spite of the doubts about the (light) scalar mesons structures
it is interesting to analyze separately the behavior of the corresponding 
quark-antiquark states for their structures \cite{NJL-scalars}.
The quark propagator will be written in the adjoint representation.
These effects will  also be computed for the ({\it diagonal}) scalar  quark-antiquark states
 that will be associated to the mesons $f_0(500)$ (or $\sigma(500)$),
 $f_0(980), \chi_{c0}, \chi_{b0}$.

Therefore in  this work  
 flavor-dependent 
coupling constants for the flavor U(5) NJL-model 
will be derived  by considering 
one loop background field method 
\cite{BFM,PRD-2014,PLB-2016}.
Their role, and also the role of  flavor-dependent cutoffs, 
will be investigated for observables  analogously to the case of $U(3)$ NJL model worked out in
 \cite{PRD-2021} by considering  degenerate up and down quark masses.
  Mixing-type interactions  $G_{i\neq j}$,
for $i,j = 0,8,15,24$ (or $G_{f_1 f_2}$ for $f_1,f_2= u,s, c,b$),  are  found to be
proportional to differences of quark effective masses,  $(M_{f_1}^* - M_{f_2}^*)^n$,
for $n=1,2$ and considerably smaller than the diagonal ones $G_{ii}$
even if mixing heavy and light quarks.
The normalization of these mixing  coupling constants  are not easily defined
being that 
they will be  mostly neglected.
Meson 
  observables usually predicted by the NJL model will be investigated:
quark-antiquark chiral condensates (including the heavy ones),
 masses and weak decay constants of quark-antiquark pseudoscalar mesons and 
quark-antiquark scalar  states 
as components of the  scalar mesons structure.
We shall adopt 
the non-covariant three-dimensional
cutoff that  presents a more uniform interpretation when
comparing light and heavy quark dynamics.
  The role of DChSB in the HQ sector,
by means the heavy quark scalar condensate, in the results will be  verified 
by switching off their contributions in the observables for some of the 
SETs of parameters.
The work is organized as follows.
In the next Section the one-loop  background field method will be applied 
to the $N_f=5$ NJL-model and 
flavor and chiral symmetry breakings  corrections to its  coupling constant ($G_0$)
will be computed in a large quark mass expansion of the 
 quark determinant.
A   quantum mixing will be  identified in the resulting couplings 
 and 
a way of implementing flavor-dependent cutoffs unambiguously will be presented.
In section  (\ref{sec:observ}), observables  will be defined
and 
in   section (\ref{sec:numerics})  numerical results will be presented.
In the final section a summary will be  presented.

\section{ Flavor dependent   NJL coupling constants }
\label{sec:Gij}

The  generating functional of  
the flavor $U (N_f=5)$   NJL  model is given by:
\begin{eqnarray} \label{GF-NJL}
Z[\bar{\eta},\eta] = \int {\cal D} [\overline\psi, \psi]
exp \left\{  i \; \left[ S_{NJL} [\bar{\psi}, \psi] + \int_x 
\left( \overline{\eta} \psi + \eta \overline{\psi} \right) \right]
\right\}
\end{eqnarray}
where
\begin{eqnarray} \label{S-NJL}
S_{NJL} [\bar{\psi}, \psi] = \int_x \left\{
\bar{\psi} \left( i \slashed{\partial} 
- m_f \right) \psi 
+ 
\frac{G_0}{2}
 \left[
(\overline{\psi} \lambda_i \psi)^2 
+ (\overline{\psi} i \gamma_5 \lambda_i \psi)^2
\right]  \right\} 
\end{eqnarray}
where 
$\int_x = \int d^4 x$, $i = 0, ..., N_f^2-1$, \ $\lambda_0=\sqrt{2/N_f} I$ and 
  $f=u,d,s,c,b$ for 
flavor in the fundamental representation.
$\eta, \bar{\eta}$ are quark sources and $m_f$ are current quark masses 
being that up and down quarks will be considered  degenerate 
\cite{PDG}:
\begin{eqnarray}
\hat{m} = diag\left(
\begin{array}{c c c c c}
m_u & m_u & m_s & m_c & m_b
\end{array} \right).
\end{eqnarray}
The quark field will be split into a quantum quark part  ($\psi_2,\bar{\psi}_2$),
 that 
and make part of quark-antiquark mesons and of the chiral condensate,
and background quark part   ($\psi, \bar{\psi}$)\cite{EPJA-2016,PRD-2019}.
At the one loop level this splitting can be done in  terms of currents
 \cite{weinberg,EPJA-2018,BFM}
 such that  chiral symmetry is preserved, and it 
 can write:
$\overline{\psi} \psi  \to (\overline{\psi} \psi) + (\overline{\psi} \psi)_2$.
The  NJL interaction  is split  into three parts: one for 
quantum field  ($I_{4}$) one for background quarks ($I_{4,bg}$) and 
one with mixed bilinears ($I_{4,coupl}$).
For the integration of 
the quantum quark component  $(\bar{\psi} \psi)_2$  
  standard auxiliary variables $S_i, P_i$ are introduced.
by means of  unity integrals that multiply the generating functional given by:
\begin{eqnarray} \label{aux-variab-1}
1 &=&  N' \; \int \; D [S_i, P_i ]
\; 
exp \;  \left[ - \frac{ i}{2 G_0 } \int_x \;  \left[ (S^i + G_0 (\bar{\psi} \lambda^i \psi)_2)^2 
+  (P^i + G_0 (\bar{\psi}  i\gamma_5 \lambda^i \psi)_2)^2
\right] \right]
,
\end{eqnarray}
where  $N'$  is
a   normalization constant.

By integrating out the  quark field 
the following one-loop effective action is obtained:  
\begin{eqnarray} \label{S_eff-log}
S_{eff} &=& 
 - i \; Tr \ln \left[ -i \left(  S^{-1}_{0,f}   + 
G_0 \lambda_i ( \bar{\psi} \lambda_i \psi 
+ i \gamma_5 \bar{\psi} \lambda_i i \gamma_5 \psi)
\right) \right]
-  \int d^4 x  \frac{1}{2 g_4} \left[ S_i^2 + P_i^2 \right]
\nonumber
\\
&+&
 \int d^4 x \left\{  \bar{\psi}\left( i \slashed{\partial} - m_f \right)\psi 
+
\frac{G_0}{2} 
\left[ (\overline{\psi} \lambda_i \psi)^2 + (\overline{\psi} i \gamma_5 
\lambda_i \psi)^2 \right]  \right\}
.
\end{eqnarray}

From the integration of   quark field, ground state equations for the auxiliary fields
are obtained by means of saddle point equations of  the effective action,
i.e. 
$\left. \frac{\partial { S}_{eff} }{\partial S_i} 
\right|_{S^i = \bar{S}^i}  =        0,
$
and
$\left. \frac{\partial { S}_{eff} }{\partial P_i} 
\right|_{S^i = \bar{S}^i, \bar{P}_i=0}  =   0$.
One has $\bar{P}^i=0$ is  a trivial  and necessary  solution:
only the scalar auxiliary fields  can be non zero due to usual charge neutrality of the vacuum
 $\bar{S}_i$ for $i=0,3,8,15,24$.
With these fields the corresponding contributions in the fundamental representation
for the 
fermions  can be written as mass terms
$M_f  \equiv  m_f + \bar{S}_f$, where the last term
arises from particular combinations of the non zero expected values of the fields $S_i$.
The resulting  usual gap equations
for  the quark flavor $f$  can be written as:
\begin{eqnarray} \label{gap-g4}
M_f - m_f  = G_0 \; Tr \; \int \frac{d^4 k}{(2 \pi)^4} {S}_{0,f} (k).
\end{eqnarray}
where the quark propagator is:
 $S_{0,f} (k) =  \left(  i \slashed{k} - M_f + i \epsilon
\right)^{-1} $.
Since the model is a low energy non-renormalizable model
these integrations will be done by choosing the three-dimensional non-covariant
 regularization scheme in terms
 ultraviolet (UV) cutoff.

\subsection{ Expansion of the determinant}

At low energies constituent quark masses are quite  large, 
so 
 we proceed by performing a large 
quark mass expansion of the determinant and by applying a zero order derivative expansion
\cite{mosel}
such that low energy  effective couplings are resolved in the very long-wavelength local limit.
For the  leading  non derivative terms we can write:
\begin{eqnarray} \label{L-det}
{\cal L}_{det}  \simeq M_3^f  \bar\psi (x)  \lambda_f \psi (x)
+ \frac{\Delta G_{ij}^s }{2}
   (\bar\psi \lambda_i \psi)  (\bar\psi \lambda_j \psi) 
+ 
 \frac{\Delta G_{ij}^{ps} }{2}
  (\bar\psi i\gamma_5 \lambda_i \psi)  (\bar\psi i \gamma_5 \lambda_j \psi) 
,
\end{eqnarray}
where the effective parameters were  defined,  at zero momentum transfer  limit and 
after a Wick rotation to Euclidean momentum space with the corresponding quark propagator
$S_{0,f}(k)$,  by:
\begin{eqnarray}
M_3^f  &=&   G_0 \;  tr_{D,F,C} \int \frac{d^4 k}{(2 \pi)^4} \;   S_{0,f} \left(k  \right) \lambda_i
,
\\  \label{Gps}
\Delta G^{ij}_{ps}  = \Delta G^{ij}_{ps} (P^2 \to 0) &=& 2 \;  G_0^2 \;  tr_{D,F,C}
\int \frac{d^4 k}{(2 \pi)^4} \; 
 S_{0,f} \ \left(k + \frac{P}{2} \right) \; \lambda_i i \gamma_5  \; S_{0,f} 
\left(k - \frac{P}{2} \right) \; \lambda_j i \gamma_5,
\\   \label{Gs}
\Delta G^{ij}_s   = \Delta G^{ij}_s  (P^2 \to 0) &=&  2 \; G_0^2 \;   tr_{D,F,C}
\int \frac{d^4 k}{(2 \pi)^4} \; 
 S_{0,f} \left(k + \frac{P}{2} \right) \; \lambda_i  \; S_{0,f} 
\left(k - \frac{P}{2} \right) \; \lambda_j , 
\end{eqnarray}
where  
$ tr_{D,F,C}$ stands for traces in Dirac, flavor and color indices.
The parameters $M_3^f$ are proportional to 
the gap equation effective masses,  but they 
renormalize differently from the gap equation \cite{PRD-2019,YuMod-2021},
 and they  will not be worked out further in this work.
The flavor-symmetric limit is recovered immediately for equal current quark masses
and reduces to previous calculations  \cite{PLB-2016,PRD-2021},
 except for the flavor group coefficients.

The difference in scalar and pseudoscalar interactions can be arranged in 
such a way to identify a sort of nearly chiral invariant term 
- except that it breaks flavor (uniformly in the scalar and pseudoscalar channels)
 - and a chiral symmetry breaking term
in 
the following 
way:
\begin{eqnarray} \label{Gps+Gsb}
I_4 &=& 
 \frac{\Delta G_{ij}^{ps} }{2} \left[
   (\bar\psi \lambda_i \psi)  (\bar\psi \lambda_j \psi) 
+ 
  (\bar\psi i\gamma_5 \lambda_i \psi)  (\bar\psi i \gamma_5 \lambda_j \psi) 
\right]
- \frac{\Delta G_{ij}^{sb} }{2} 
   (\bar\psi \lambda_i \psi)  (\bar\psi \lambda_j \psi) 
,
\end{eqnarray}
where  $\Delta G_{ij}^{ps}$ reduces  to a chirally symmetric term in the limit of 
degenerate quark masses
and 
the coupling constant for the symmetry breaking term,  in  the scalar sector, is given by:
\begin{eqnarray} \label{Gsb}
\Delta G_{ij}^{sb}  &=& 
4  \;  G_0^2 \; tr_{D,F,C}
 \int_k \tilde{S}_{0,f} \left(k + \frac{P}{2} \right) \;  \lambda_i
 \; \tilde{S}_{0,f} 
\left(k - \frac{P}{2} 
\right)  \; \lambda_j
\;
\end{eqnarray}
where $\int_k = \int \frac{ d^4 k}{(2\pi)^4}$ and, in Euclidean momentum space:
\begin{eqnarray} 
\tilde{S}_{0,f} (k) &=& 
\frac{ M_f }{ k^2 + {M_f}^2}  .
\end{eqnarray}
Although this way of separating chirally symmetric 
and symmetry breaking terms is somewhat arbitrary 
it is very interesting and suitable because 
it leaves the symmetry breaking term directly proportional to the 
quark effective mass.
Note that $\Delta G_{ij}^{sb}$ is  directly proportional to $M_f$ and therefore
is manifestation of the explicit and dynamical chiral symmetry breakings.
The limit of chiral and flavor symmetric interaction
is obtained with   $\Delta G_{ij}^{sb} \to 0$.
These traces in flavor indices were computed by considering the quark 
(effective) propagators in the adjoint representation.  
These propagators are parameterized in 
Appendix (\ref{app:quarkprop}).
The differences between the scalar and pseudoscalar corrections have been investigated
for the flavor  SU(3) model in the vacuum and under magnetic fields and it was 
found that, to describe light mesons phenomenology, it is suitable  to neglect either 
the scalar \cite{PRD-2021,JPG-2022} 
or the pseudoscalar flavor-dependent  interactions \cite{PRD-2022b} respectively
in the vacuum and at finite magnetic field.
 In the present work, the pseudoscalar (scalar) interaction
will be considered for the pseudoscalar (scalar) mesons structures.

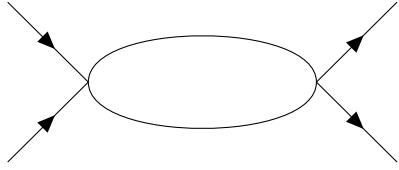
\begin{figure}[ht!]
	\begin{center}
		\begin{tikzpicture}
		\begin{feynman}
		\vertex (a);
		\vertex [right=of a] (c);
	    \vertex [right=of c] (b);
		\vertex [above left=of a] (t1);
		\vertex [below left=of a] (t2);
	    \vertex [above right=of b] (s1);
		\vertex [below right=of b] (s2);
		
		\diagram* {
			(t2) -- [fermion] (a),
	        (t1) -- [fermion] (a)
			-- [half left, looseness=0.7, edge] (b)
			-- [half left, looseness=0.7, edge] (a),
			(b) -- [fermion] (s1),
	        (b) -- [fermion] (s2),
		};
		\end{feynman}
		\end{tikzpicture}
	\end{center}
\caption{ \label{diagrams-2}
	\small
	This is the 4-leg one-loop diagram calculated with
 the second order term of the determinant expansion.
	Numerical values are extracted for zero momentum transfer.
}
\end{figure}

After the inclusion of the coupling constants obtained from the quark 
determinant, one obtains:
\begin{eqnarray}  \label{Gij-s-ps}
G_{ij}^{s,ps} = G_0 \delta_{ij} + \Delta G_{ij}^{s,ps},
\end{eqnarray}
being that a nearly chiral invariant limit can be obtained only
by renormalizing in the point $\Delta G^{sb}_{ij} = 0$.
The  eqs. above (\ref{Gps+Gsb})
contain  corrections to a NJL coupling constant $G_0$ being that the 
flavor dependencies are due to the non degenerate quark masses.
To assure that the flavor dependent interactions contribute strictly for the original model with $G_0$,
whose value  is somehow arbitrary in the NJL model,
 one chooses to re-normalize the resulting 
total coupling constant.
Therefore the following re-normalization
will be considered:   
\begin{eqnarray} \label{Gnorm}
G_{ij}^{n,ps} &=& G_0 \frac{ G_{ij}^{ps} }{ G_{11}^{ps} },
\;\;\;\;\;\;\;\; \mbox{such that } \;\;\; G_{11}^{n,s/ps} = 10 \; \mbox{GeV}^{-2},
\\ 
\label{Gineqj}
G_{ij}^{n,s} &=& G_0  \frac{ G_{ij}^{s} }{ G_{11}^{s} },
\end{eqnarray}
where the superscript $^n$ indicate the normalized coupling constants.
From here on, we'll be dealing with the normalized coupling constants
and this superscript will be omitted.
The advantage of this normalization is that, by starting from an
 effective coupling constant of reference, $G_0$, one avoids
artificial excessive increase of the overall  and resulting coupling constants
which can happen by adding quantum fluctuations. 
Afterall
the initial numerical value of $G_0$ can be attributed to 
one (non perturbative) gluon exchange processes 
and therefore it can be re-arranged to incorporate
corrections and to fit observables.
The choice of renormalizing all the channels $(ij)$ with respect to $i,j=1$
(and consequently to $G_{22}=G_{11}$) was taken because it defines the 
charged pion,  that was chosen to be kept without further
changes. The modifications on the other meson states can be assessed accordingly.
The normalization for the $G_{i \neq j}$ is less obvious and they may be different from the
above one
 since they are 
exclusively due to the quantum fluctuations leading to  an ambiguity in eq. (\ref{Gineqj}).
Being considerably smaller than
the diagonal ones $G_{ii}$, they will not be taken into account  in the gap 
equations and the bound state equations.
Note that, $G_{i j} = G_{j i}$ as a result from CPT and electromagnetic U(1) 
symmetries.
In Fig.  (\ref{diagrams-2})
 the Feynman diagram for the one loop
  second order terms of the expansion is drawn.


\FloatBarrier

A resulting NJL-model Lagrangian,
with the normalized coupling constants, can be written as:
\begin{eqnarray} \label{S-NJL-Gij}
{\cal L}_{NJL} &=& 
\bar{\psi} \left( i \slashed{\partial} 
- m_f \right) \psi 
+ 
\frac{G_{ij}^s}{2}
(\overline{\psi} \lambda_i \psi) (\overline{\psi} \lambda_j \psi) 
+ 
\frac{G_{ij}^{ps}}{2}
(\overline{\psi} i \gamma_5 \lambda_i \psi) (\overline{\psi} i \gamma_5 \lambda_j \psi),
\end{eqnarray}
where the contributions $\Delta G_{ij}^{sb}$ will be neglected.
 Some works argued that the light (l) and heavy (h) quark sectors  in the NJL model
should feel different interactions.
So, by neglecting the mixing interactions
and by considering $G_{ii} = G_{ll}$ for $i=1,...,8$, $G_{ii}=G_{hl}$ for $i=9,...,14,16, ...,21$
and $G_{ii} = G_{hh}$ for $i=22,23$ in the SU(5) limit of the above 
Lagrangian,  one nearly recovers the basis of other works, for example 
\cite{npb-ebert-etal-1995}.

\subsection{ Representations of the flavor group and quantum mixing}

The gap equations of the model (\ref{S-NJL-Gij}) have to be calculated.
 By neglecting all  mixing interactions, $G_{i\neq j}$ and $G_{f \neq g}$,
the new (corrected) gap equations with the flavor-dependent coupling constants
 can be written as:
 \begin{eqnarray} \label{gap-gff}
M_f^* - m_f  = G_{ff} \; Tr_{D,C,F} \; \int \frac{d^4 k}{(2 \pi)^4} {S}_{0,f} (k),
\end{eqnarray}
where  
 the corrected quark propagator (diagonal matrix) is
${S}_{0,f} (k) = (\slashed{k} - M^*_f + i\epsilon)^{-1}$.
The flavor-symmetric limit is recovered immediately for equal current quark masses.
For the interactions of color singlet and flavor singlet currents,
these coupling constants $G_{ff}$ can be defined 
as:
\begin{eqnarray}   \label{Gf1f2}
G_{f_1 f_2} (\bar{\psi} \psi)_{f_1} (\bar{\psi} \psi)_{f_2}
= 2  G_{ij} (\bar{\psi} \lambda_i \psi ) ( \bar{\psi} \lambda_j \psi ),
\;\;\;\;\;\;\;
f_1, f_2 = u,d,s,c,b.
\end{eqnarray}
Since the mixing interactions $G_{f_1f_2}$ are considerably smaller and will be neglected,
 the diagonal ones $G_{ff}$ are given by:
\begin{eqnarray} \label{Guu}
G_{uu} &=& \frac{2}{5}G_{0,0} + G_{3,3} + \frac{4}{\sqrt{30}}G_{0,8} + \frac{2}{\sqrt{15}}G_{0,15} + \frac{2}{5}G_{0,24} + \frac{1}{3}G_{8,8} \nonumber \\
& & + \frac{2}{3\sqrt{2}}G_{8,15} + \frac{2}{\sqrt{30}}G_{8,24} + \frac{1}{6}G_{15,15} + \frac{1}{\sqrt{15}}G_{15,24} + \frac{1}{10}G_{24,24}, \label{Guu}
\\
G_{ss} &=& \frac{2}{5}G_{0,0} - \frac{8}{\sqrt{30}}G_{0,8} + \frac{2}{\sqrt{15}}G_{0,15} + \frac{2}{5}G_{0,24} + \frac{4}{3}G_{8,8} \nonumber \\
& & - \frac{4}{3\sqrt{2}}G_{8,15} - \frac{4}{\sqrt{30}}G_{8,24} + \frac{1}{6}G_{15,15} + \frac{1}{\sqrt{15}}G_{15,24} + \frac{1}{10}G_{24,24}, \label{Gss}
\\
G_{cc} &=& \frac{2}{5}G_{0,0} - 2 \sqrt{\frac{3}{5}}G_{0,15} + \frac{2}{5}G_{0,24} + \frac{3}{2}G_{15,15} - \sqrt{\frac{3}{5}}G_{15,24} + \frac{1}{10}G_{24,24}, \label{Gcc}
\\
G_{bb} &=& \frac{2}{5}G_{0,0} - \frac{8}{5}G_{0,24} + \frac{8}{5}G_{24,24}. \label{Gbb}
\end{eqnarray}
When inverting these relations it is seen that even the interactions among heavy quarks
can contribute to the dynamics of the light ones and
they will contribute to the light  quark-antiquark mesons states
- defined in the adjoint representation -  with different intensities.
This is a quantum mixing \cite{sakurai} for the
punctual 4-point interaction of the model representing 
internal (quark-antiquark) degrees of freedom of the quark model.
These
 flavor-dependent coupling constants can be associated 
to the (three-channel) two-dressed gluon exchange process \cite{PRD-2021,JPG-2022}.

 \subsection{ Flavor dependent cutoffs }
\label{sec:cutoff}

It is usually argued  that the UV cutoff 
used in effective models for heavy and light quarks must be different because
 light and heavy quarks feel differently strong interactions 
and they should be subject to different energy scales due to the very different quark
masses \cite{SBK,mexicanos-2019,bashir-etal,craig-etal}. 
However in the polarization tensor the contributions for heavy-light 
quark pairs it is not obvious how to implement such different
flavor-dependent $\Lambda_f$ since the 
individual internal three-momenta of the heavy and light quarks
are subject to different energy scales.
Below a systematic way of implementing is developed by 
keeping conservation laws safe.

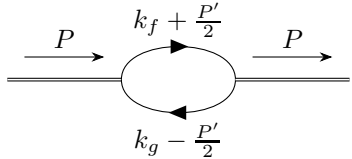
\begin{figure}[ht!]
\begin{center}  
	\begin{tikzpicture}
	\begin{feynman}
	\vertex (a);
	\vertex [right=of a] (b);
	\vertex [right=of b] (c);
	\vertex [right=of c] (d);
	
	\diagram* {
		(a) -- [double, momentum=\(P\)] (b)
		-- [fermion, half left, looseness=1.0, edge label=\(k_f+\frac{P'}{2}\)] (c)
		-- [fermion, half left, looseness=1.0, edge label=\(k_g-\frac{P'}{2}\)] (b),
		(c) -- [double, momentum=\(P\)] (d),
	};
	\end{feynman}
	\end{tikzpicture}
\end{center}
	\caption{ \label{fig:diagrams-3}
		\small
		Momentum dependence of the one-loop meson structure.
The internal lines have  $\vec{P'}$ not necessarily equal to $\vec{P}$ to 
allow for conservation of momenta in the case of  non degenerate cutoffs.
	}
\end{figure} 

\FloatBarrier

Consider the  diagram (\ref{fig:diagrams-3}) for the meson structure
with external  meson four-momentum    $P$ 
and internal momentum $k$.
To implement the different cutoffs 
the internal momentum will be make different into $k_f$ and $k_g$ 
for two different quarks.
A rescaling of the (internal) three dimensional momentum can be done such as:
\begin{eqnarray}
k_f = ( k_0, \delta_f \vec{k} ),
\;\;\;\;\;\; k_g = ( k_0, \delta_g \vec{k} ),
\;\;\;\;\;\;  \delta_f = \Lambda_f / \Lambda .
\end{eqnarray}
Where  $\delta_f$ or $\delta_g$ are constants  such that the maximum momentum (cutoff)
will be  $\Lambda_f = \delta_f \Lambda$ and $\Lambda_g = \delta_g \Lambda$.

The  conservation of energy-momentum for 
the rest frame of the meson  can be re-written as:
\begin{eqnarray}
(P_0 , 0 ) &=& 
( k_0, \delta_f \vec{k} ) + \frac{1}{2} (P_0 , 
\vec{P}' - \vec{P}) - ( k_0, \delta_g \vec{k} )
+ \frac{1}{2} ( P_0 , \vec{P}' - \vec{P}) 
\end{eqnarray}
With these parameterizations, the following relations 
for the heavy-light equations are obtained:
$\vec{P}' - \vec{P}  = (\delta_g - \delta_f ) \vec{k} $.
and the relation dispersion of each internal line for 
a polarization tensor with quarks of different flavor can be rewritten as:
\begin{eqnarray} \label{E-C}
E_f^2 &=& (C_{\delta, fg} \vec{k} )^2 + M_f^2, 
\;\;\;\;\; C_{\delta, fg} = \frac{ \delta_f + \delta_g}{2} .
\end{eqnarray}
These  phenomenological parameters, $\delta_f$,
may be expected to  make possible further improvements
in the fitting procedure.
It will be shown however, that their role may not really very important
since results do not change considerably if $\delta_f=1$.

\section{ Observables}
\label{sec:observ}

In 
this section some observables are presented
for the flavor-dependent coupling constants.
The corresponding equations are trivially obtained for the
coupling constant of reference $G_0$.
They will also be used to make possible 
an assessment of the effect of the $G_{ij}$.

\subsection{ Pseudoscalar and scalar mesons spectrum}

In the standard treatment of meson as quark-antiquark bound state in the NJL model
the
 Bethe Salpeter equation, or bound state equation, (BSE)
has a constant amplitude
 in the Born 
approximation \cite{NJL,klevansky,vogl-weise,hatsuda-etal,klimt-etal}.
It can be obtained
from   the condition of (real)
pole for the propagator of the auxiliary fields  describing the quark-antiquark state in the 
corresponding flavor-Dirac channel.
It can be written as:
\begin{eqnarray} \label{BSE-1}
1 - 2 \; G_{ij} \; Tr_{DCF} \; \left(
\int_k \lambda_i \; \Gamma  \;  S_{0,f_1}
( k + P/2)  \; \lambda_j \; \Gamma \;  S_{0,f_2} (k - P/2) \right)  = 0,
\end{eqnarray}
where $\Gamma$ stands for the Dirac pseudoscalar and scalar channels operators, respectively
$\Gamma = i \gamma_5$ and $I$.
This momentum  integral corresponds 
to the polarization tensor with quadratic and logarithmic UV divergences
which are most appropriated  regularized by the same UV cutoff of the gap equations.
The indices in both adjoint and fundamental representation are needed in this equation
and they are tied by the internal structures of the bound meson states.
The basic meson structures considered  are presented in Table (\ref{table:comp})
for the pseudoscalar and scalars.
However, as seen above, the non degeneracy of quark masses lead to 
mixing terms when calculated the traces in flavor indices
if all the matrices.
The terms resulting  from the complete polarization tensor 
  are presented in Table (\ref{table:Gij}).
These mixing terms plays an important role for the 
pseudoscalars $\eta,\eta',\eta_c,\eta_b$
that will be calculated as $P_0, P_8, P_{15}$ and $P_{24}$.
The corresponding scalar mesons
 (built as  $S_0, S_8, S_{15}$ and $S_{24}$)
are not necessarily quark-antiquark states, therefore
we label the corresponding quark-antiquark states.
These states might be part of the structure of the mesons
$\sigma(500),f_0(980),\chi_{0,c}$ and $\chi_{0,b}$.
These states ($\phi_i= P_i, S_i$) are defined as:
\begin{eqnarray} \label{mesons-0-8-etc}
 \phi_0 &=& \sqrt{\frac{{2}}{5}} ( \bar{u} u + \bar{d} d +  \bar{s} s + \bar{c} c + \bar{b} b),
\nonumber
\\
 \phi_8  &=& \frac{1}{\sqrt{3}} ( \bar{u} u + \bar{d} d -2  \bar{s} s )
\nonumber
\\
 \phi_{15}  &=& \frac{1}{\sqrt{6} } ( \bar{u} u + \bar{d} d +  \bar{s} s -3 \bar{c} c ),
\nonumber
\\
 \phi_{24}  &=& \frac{1}{\sqrt{10}}
 ( \bar{u} u + \bar{d} d +  \bar{s} s + \bar{c} c - 4 \bar{b} b),
\end{eqnarray}
The solutions for the BSE for scalar mesons, Eq. (\ref{BSE-1}) for $\Gamma = I$,
present some subtlety because the meson masses are above the 
threshold of the unphysical  decay into the pair of quark-antiquark.
A particularly convenient way to deal with this problem is to 
introduce an infrared cutoff ($\Lambda_{IR}$) \cite{cutoff-IR} as discussed below.

\begin{table}[ht]
	\caption{
		\small  Composition $(\bar{\psi}\lambda_i\psi)$ of pseudoscalar  (scalar) mesons states
for the case where up and down quarks are degenerate, $l=u,d$.} 
	\centering 
	\begin{tabular}{c  c c c c c c c c c c c } 
		\hline 
		meson  & & & & & & & & & & & 
		\\
		P (S) & $\pi (a_0)$ & $K (K_0^*)$ 
& $D (D_0^*)$ & $D_s (D_{s0}^*)$ & $B (B_0^*)$ & $B_s (B_{s0}^*)$ & $B_c (B_{c0}^*)$
 & $\eta' (S_0)$ & $\eta (S_8)$ 
& $\eta_c (S_{15})$ & $ \eta_b (S_{24})$  
		\\
		\hline
		\\
		$i$ & 1-3 & 4-7 & 9-12 & 13-14 & 16-19 & 20-21 & 22-23 & 0 & 8 & 15 & 24
		\\
		\\ 
		$\bar{f}f$ & $\bar{l}l$ & $\bar{s}l$ & $\bar{l}c$ & $\bar{s}c$ & $\bar{b}l$ & $\bar{b}s$ & $\bar{b}c$ & $P_0 (S_0)$  & $P_8 (S_8)$ & $P_{15} (S_{15})$ & $P_{24} (S_{24})$
		\\ 
		\hline 
	\end{tabular}
	\label{table:comp} 
\end{table}

In the present model, although the effective interaction due to the  axial anomaly 
was not included, the flavor-dependent coupling constants
are responsible for 
different types of  mixings in the spectrum.
The first ones were the quantum mixing,  identified above, and the 
mixing interactions $G_{f_1\neq f_2}, G_{i\neq j}$.
Another source of mixing appears in the calculation of the 
polarization tensor in Eq. (\ref{BSE-1}) for the meson structure.
In spite of the fact that the commuting generators
 $\lambda_0,\lambda_8,\lambda_{15}$ and $\lambda_{24}$
are diagonal,  
 the 
integrals in the fundamental representation above
 contain mixing terms ${\cal J}_{us}, {\cal J}_{uc}, {\cal J}_{ub}, {\cal J}_{sc}, {\cal J}_{sb},
{\cal J}_{cb}$
because of the quark masses non degeneracy.
These mixing-parts of the polarization tensor will provide important contributions
for the mesons (S or P) built up with the diagonal generators.
To make clear the interpretation of the quark-antiquark pairs that contribute
for each meson state,  in Table (\ref{table:Gij})
the quark-antiquark content of each flavor contribution $\Delta G_{ij}^{s,ps}$ is presented
by means of the following integrals:
\begin{eqnarray} \label{polariz-rep}
{\mathcal J}_{f_1 f_2} (P^2) &=& 
 i  N_c \; Tr_D \; 
 \int_k S_{0,f_1} \left(k + \frac{P}{2} \right) \; S_{0,f_2} 
\left(k - \frac{P}{2} \right) \;   ,
\end{eqnarray}
for both $\Gamma = I , i \gamma_5$.
With these quantities ${\cal J}_{f_1 f_2}$
the corresponding polarization tensors $\Pi_{ij}(P^2)$ are formed.

\begin{table}[h]
\centering
	\begin{tabular}{c|c} \hline \centering
		& $\Pi_{ij}$ \\ \hline\hline
		$i=j=$ 1-3 & $\mathcal{J}_{uu}$\\
		$i=j=$ 4-7 & $\mathcal{J}_{us}$\\ 
		$i=j=$ 8 & $\frac{1}{3} \left(\mathcal{J}_{uu} + 2\mathcal{J}_{ss}\right)$\\
		$i=j=$ 9-12 & $\mathcal{J}_{uc}$\\
		$i=j=$ 13-14 & $\mathcal{J}_{sc}$\\
		$i=j=$ 15 & $\frac{1}{12} \left(2\mathcal{J}_{uu} + \mathcal{J}_{ss} + 9\mathcal{J}_{cc}\right)$\\
		$i=j=$ 16-19 & $\mathcal{J}_{ub}$\\
		$i=j=$ 20-21 & $\mathcal{J}_{sb}$\\
		$i=j=$ 22-23 & $\mathcal{J}_{cb}$\\
		$i=j=$ 24 & $\frac{1}{20} \left(2\mathcal{J}_{uu} + \mathcal{J}_{ss} + \mathcal{J}_{cc} + 16\mathcal{J}_{bb}\right)$\\
		$i=j=$ 0 & $\frac{1}{5} \left(2\mathcal{J}_{uu} + \mathcal{J}_{ss} + \mathcal{J}_{cc} + \mathcal{J}_{bb}\right)$\\
		$i,j=$ 0,8 & $\frac{2}{\sqrt{30}} (\mathcal{J}_{uu} - \mathcal{J}_{ss})$ \\
		$i,j=$ 0,15 & $\frac{1}{2\sqrt{15}} (2\mathcal{J}_{uu} + \mathcal{J}_{ss} - 3\mathcal{J}_{cc})$ \\
		$i,j=$ 0,24 & $\frac{1}{10} (2\mathcal{J}_{uu} + \mathcal{J}_{ss} + \mathcal{J}_{cc} - 4\mathcal{J}_{bb})$ \\
		$i,j=$ 8,15 & $\frac{1}{3\sqrt{2}} (\mathcal{J}_{uu} - \mathcal{J}_{ss})$ \\
		$i,j=$ 8,24 & $\frac{1}{\sqrt{30}} (\mathcal{J}_{uu} - \mathcal{J}_{ss})$ \\
		$i,j=$ 15,24 & $\frac{1}{4\sqrt{15}} (2\mathcal{J}_{uu} + \mathcal{J}_{ss} - 3\mathcal{J}_{cc})$ \\
		\hline\hline
	\end{tabular}
	\caption{\small 
 Components of the polarization for $u=d$ of eq. (\ref{polariz-rep}).
 }
\label{table:Gij}  
\end{table}

The on-shell 
pseudoscalar meson-quark coupling constants were calculated as the residue of the pole 
of the T-matrix for the bound state  amplitude with eq. (\ref{BSE-1}).
It can be written   as:
\begin{eqnarray} \label{Gpsqq}
g_{psqq} =  \left( \frac{ \partial \Pi_{ij} (P^2)
 }{\partial P_0^2 } \right)^{-2}_{(P_0, \vec{P})\equiv M_{ps}},
\end{eqnarray}
where the flavor indices are tied with the quantum numbers of the 
meson $P$ as shown above.

The  charged pseudoscalar mesons weak decay
 constant were also calculated  at the one loop level \cite{klevansky,vogl-weise}
being given by:
\begin{eqnarray}
f_{ps} =  \frac{N_c  \; g_{psqq}}{4}\; \int \frac{ d^4 k}{(2 \pi)^4}
 Tr_{F,D} \left[ \gamma_\mu \gamma_5 
\lambda_i \; S_{f_1} (k + P/2) \lambda_j S_{f_2} (k - P/2) \right]_{(P_0,\vec{P})=M_{ps}},
\end{eqnarray}
where $f_1,f_2$ correspond to the quark/antiquark of the  meson 
and $i,j$ are the associated flavor indices as discussed above.
The same mixings in the BSE for the diagonal states, $\phi_0,\phi_8,\phi_{15},\phi_{24}$
appear in this decay constants.

Finally, 
the  chiral quark-antiquark scalar  condensates, as the order parameters of 
DChSB,
can be written as:
\begin{eqnarray}
\left< \bar{q}q \right>_f  \; =  \; Tr \; [ \; S_{0,f} (k=0) \; ].
\end{eqnarray}
Therefore their  values can be directly extracted from the gap equations.
 Their values will be obtained
 for the two choices of coupling constants: $G_0$ and $G_{ff}$
 and for the two different sets of 
parameters, $\Lambda_0$ and $\Lambda_f$. 
This way
the effects of the flavor dependence of the coupling constants 
and  cutoffs can be assessed.

\section{Numerical results}
\label{sec:numerics}

First of all, the fitting of the parameters of the model
($m_u,m_s,m_c,m_b$ and $\Lambda$) for a given $G_0=10$ GeV$^{-2}$
was done. For this, 
the gap equations and BSE were calculated to reproduce the masses of 
  seven neutral pseudoscalar mesons
- $\pi, K, D, D_s, B, B_s, B_c$ -  as fitting observables. Two SETs of parameters were 
fixed in this way, which  will 
be denoted SET 1 and SET 3. 
A second fitting procedure was adopted by considering
the $G_{ij}$ and $G_{ff}$ in the BSE and gap equations - SETs 2 and 4.
The four $\eta's$ particles, however,   were not considered for the 
fitting procedure so that their masses will be obtained as predictions.
After that, the gap equations of SETs 1 and 3 were recalculated with $G_{ff}$ 
and BSE for    the coupling constants $G_{ij}$ 
to make possible an assessment of their effects.
Similarly, for SETs 2 and 4, in a second stage, the coupling constants $G_{ij}$ and $G_{ff}$ were
reduced to $G_0$ to analyze the effects of the flavor dependent coupling constants in an "opposite
procedure".
A perturbative approach was envisaged, i.e.  without the self-consistency
in calculating $M^*_f$ and $G_{ij}$.
For the flavor U(3)  light sector, the self-consistent calculation  \cite{JPG-2022} may 
yield results 
similar to the perturbative calculation \cite{PRD-2021}
 with a   renormalization of coupling constants in the BSE.
The cutoffs were unique, i.e. flavor independent, for SETs 1 and 2 - $\delta_f=1$ -, while
the flavor dependent cutoffs were considered for SETs
3 and 4.
By analyzing the best values of the parameters of the model, by including $\delta_f$,
it was found that  the light quark sector should  not be modified with these extra parameters,
i.e. $\delta_f=1$ for  $f=u,s$, 
 whereas
 the  heavy quark propagators were found to undergo  small changes 
due to those modifications.
Therefore, at the end, only two cutoffs will be considered: one for the light quarks and one for the 
heavy quarks.
 In total there will be five or six parameters of the model (for one or  two cutoffs)
 and seven fitting observables
(non diagonal pseudoscalar meson masses)  for a fixed $G_0$.

\subsection{ Sets of parameters of the model}

The  SETS of parameters  were found by fitting the (first) seven pseudoscalar meson masses 
of Table (\ref{table:comp}).
As discussed above the following different types of fitting-calculations were considered:
\begin{itemize}
\item SET 1: 
 with
  non covariant  flavor-independent
cutoff ($\Lambda_0$)
and 
gap equations and  BSE calculated with $G_0$.
\item SET 2: 
 with 
non covariant  flavor-independent
cutoff ($\Lambda_0$) and
gap equations and BSE calculated  with  $G_{ij}^{s,ps}$
and $G_{ff}^{s,ps}$.
\item SET 3: with non covariant  flavor-dependent
cutoff  ($\Lambda_f$) and   BSE calculated with $G_0$.
\item SET 4: with  non covariant  flavor-dependent
cutoff  ($\Lambda_f$) 
and BSE calculated with $G_{ij}^{s},G_{ij}^{ps}$.
\end{itemize}

In Table  (\ref{tab:parameters-gap})
the 
parameters of the model, current quark masses and UV cutoff,
for each of these fitting procedures  
    are shown 
together with  
the resulting  effective masses for the calculations with $G_0$ and with $G_{ff}$,
respectively
 $M_f$ and $M_f^*$.
Up and down quarks are taken degenerate, so some  
mass splittings between charged and neutral 
mesons states cannot show up.
The   current quark masses were kept  within error bars of the  values from 
Particle Data Book (PDG) \cite{PDG},
although this should not be necessarily a strong constraint to parameters in the NJL model.
Their use and effect in gap equations have 
 a  non-linear dependence on  $\left< \bar{q}q\right>$, and this may involve
a slightly  different meaning than those extracted in PDG.
The parameter $\xi_f = m_f/M_f$ provides the ratio of the current quark masses to the effective quark masses \cite{craig-etal}.
It measures the relative contribution of the  chiral condensate for the corresponding 
effective quark mass that is $(1 - \xi_f)$.
Two values for $\xi_f$ were provided: when calculated with $M_f$ from 
gap equations with $G_0$ and from gap equations with $G_{ff}^s$.
Note that, the values of the effective masses calculated with 
$G_{ff}^s$ or $G_{ff}^{ps}$ are usually  different. 
SETs of parameters 3 and 4 are the only ones in which the 
cutoffs were made flavor-dependent, therefore in SETs 1 and 2
$\delta_f=1$ naturally.
Curiously, one needed basically the same value for $\delta_f$ (or conversely $\Lambda_f$)
for the set of light quarks and the same $\delta_f$
 for the set of heavy quarks.

\begin{table}[ht]
	\caption{ \small 
		Sets of parameters and 
solutions of the gap equations for
standard NJL with $G_0$ and for flavor-dependent NJL coupling constants
$G_{ff}$.}
	\centering \label{tab:parameters-gap}
	\begin{tabular}{| c | c | c c c c | }  \hline \hline  
		Set 			& $f$ 								& $u$ 	& $s$ 	& $c$  	& $b$  \\ \hline \hline 
		1   			& $m_f$ (MeV)       				& 7.0 	& 145 	& 1300	& 4650
		\\
		    			& $\Lambda_f=\Lambda_0$ (MeV)		& 510 	& 510 	& 510 	& 510 
 		\\
		($G_0$)			& $M_f$ (MeV)       				& 395 	& 596 	& 1826 	& 5186 
		\\
		($G_{ff}^{ps}$)	& $M^*_f$ (MeV)     				& 395 	& 571 	& 1752 	& 5092
		\\
		($G_{ff}^s$)	& $M^*_f$ (MeV)     				& 395 	& 566 	& 1773 	& 5132
		\\
    					& $\xi_f$ ($G_0$)					& 0.018 & 0.243 & 0.712 & 0.897
    	\\
    					& $\xi_f$ ($G_{ff}^{ps}$)			& 0.018 & 0.254 & 0.742 & 0.913
		\\
    					& $\xi_f$ ($G_{ff}^s$)				& 0.018 & 0.256 & 0.733 & 0.906
 		\\ \hline 
		2   			& $m_f$ (MeV)       				& 7.0 	& 145 	& 1300	& 4650
		\\
						& $\Lambda_f=\Lambda_0$ (MeV)		& 520 	& 520 	& 520 	& 520 
		\\
		($G_0$)			& $M_f$ (MeV)       				& 430 	& 628 	& 1857 	& 5218 
		\\
		($G_{ff}^{ps}$)	& $M^*_f$ (MeV)     				& 430 	& 601 	& 1780 	& 5118
		\\
		($G_{ff}^s$)	& $M^*_f$ (MeV)     				& 430 	& 598 	& 1805 	& 5164
		\\
						& $\xi_f$ 	  ($G_0$)				& 0.016 & 0.231 & 0.700 & 0.891
		\\
						& $\xi_f$ ($G_{ff}^{ps}$)			& 0.016 & 0.241 & 0.730 & 0.909 
		\\
    					& $\xi_f$ ($G_{ff}^s$)				& 0.016 & 0.242 & 0.720 & 0.900 
		\\ \hline 
		3   			& $m_f$ (MeV)       				& 7.0 			& 150 		   & 1350		& 4700
		\\
						& $\Lambda_f$ (MeV) [$\delta_f$]
															& 530 [1.00]	& 530 [1.00]   & 510 [0.96]	& 510 [0.96]
		\\
		($G_0$)			& $M_f$ (MeV)       				& 467 			& 667 		   & 1876 		& 5236 
		\\
		($G_{ff}^{ps}$)	& $M^*_f$ (MeV)     				& 467 			& 640 		   & 1805 		& 5142
		\\
		($G_{ff}^s$)	& $M^*_f$ (MeV)     				& 467 			& 638 		   & 1830 		& 5188
		\\
						& $\xi_f$ 	 ($G_0$)				& 0.015 		& 0.225 	   & 0.720 		& 0.898
		\\
						& $\xi_f$ ($G_{ff}^{ps}$)			& 0.015 		& 0.234 	   & 0.748 		& 0.914
		\\
    					& $\xi_f$ ($G_{ff}^s$)				& 0.015 		& 0.235 	   & 0.738 		& 0.906
		\\ \hline 
		4   			& $m_f$ (MeV)       				& 6.5 			& 140 		   & 1240		& 4580
		\\
						& $\Lambda_f$ (MeV) [$\delta_f$]
															& 530 [1.00]	& 530 [1.00]   & 540 [1.02] & 540 [1.02] 
		\\
		($G_0$)			& $M_f$ (MeV)       				& 466 			& 654 		   & 1863 		& 5216 
		\\
		($G_{ff}^{ps}$)	& $M^*_f$ (MeV)     				& 466 			& 628 		   & 1778 		& 5105
		\\
		($G_{ff}^s$)	& $M^*_f$ (MeV)     				& 466 			& 627 		   & 1808 		& 5159
		\\
						& $\xi_f$ 	 ($G_0$)				& 0.014 		& 0.214 	   & 0.666 		& 0.878
		\\
						& $\xi_f$ ($G_{ff}^{ps}$)			& 0.014 		& 0.223 	   & 0.697 		& 0.897
		\\
    					& $\xi_f$ ($G_{ff}^s$)				& 0.014 		& 0.223 	   & 0.686 		& 0.888
		\\ \hline \hline 
	\end{tabular}
\end{table}

\subsection{ Coupling constants}

In Table (\ref{table:Gps}) the resulting normalized flavor-dependent coupling constants
calculated with Eqs. (\ref{Gnorm},\ref{Gij-s-ps}) are shown for the  SETs of parameters
for all $i,j$ channels.
Before normalization, 
the corrections to the diagonal coupling constants $\Delta G_{i i}$
are, at most, nearly of the order of  $0.25 \times G_0$  
in agreement with previous estimates for polarization effect for light quarks 
\cite{PRD-2014,PLB-2016}.
In the last line the absolute value  of the  total $G_{11}^s, G_{11}^{ps}$ 
 that were used for the normalization are exhibited.
In average, the scalar (pseudoscalar) channel presents smaller  coupling constants - normalized or not -
for the light (heavy) quark sector, i.e. $i=j=1,...,8$ ($i=j=9,...,24$),
with few exceptions.
Note that the heavy flavor coupling constants become smaller than the light flavor ones.
Also the mixing interactions $G_{i\neq j}$ are considerably smaller since they 
are proportional to $(M_{f_1} - M_{f_2})^n$ for $n=1,2$.
Among the mixing coupling constants, $G_{0,15}$ and $G_{0,24}$ are the largest ones
and the mixing $G_{08}$, that contribute for the $\eta-\eta'$ mixing
\cite{PRD-2021,JPG-2022} is not so large, although 
it is not among the smallest.
When comparing the different diagonal channels $G_{ii}$
it can be seen that there are groups of couplings that are closer to each other:
$G_{ii}$ for $i=4-8$ (slightly different from $i=1-3$ - the pion channel),
$G_{ii}$ for $i=9-15$ and  $G_{ii}$ for $i=16-24$.
This is quite in agreement with the idea of separating 
light, heavy-light and heavy  quark (or quark-antiquark) interactions
\cite{npb-ebert-etal-1995}.

In Table (\ref{table:Gffps})
the normalized coupling constants 
in the fundamental representation
$G_{ff}^{ps,n}$ and $G_{ff}^{s,n}$ 
 are presented
for the  normalized $G_{ij}^{n,ps}$ and $G_{ij}^{n,s}$.

\begin{table}[ht]
	\caption{
		\small
		Flavor-dependent
coupling constants $G_{ij}^{n,ps}$ and $G_{ij}^{n,s}$ according to eqs. 
(\ref{Gij-s-ps}) and 
(\ref{Gnorm}).
Corresponding current and 
effective masses and UV cutoffs from Table  (\ref{tab:parameters-gap})
		and  $G_0=$ 10.0\,GeV$^{-2}$.
	} 
	\label{table:Gps}
	\centering
 	\begin{tabular}{| l ||c c c c||c c c c |} \hline \centering
		 & \multicolumn{4}{c||}{$G_{ij}^{n,ps}$ (GeV$^{-2}$)} 	& \multicolumn{4}{c|}{$G_{ij}^{n,s}$ (GeV$^{-2}$)} 
		\\
		SET 		   & 1	   & 2	   & 3	   & 4	   			& 1	   & 2	   & 3     & 4
		\\ \hline
		$i=j = $ 1-3   & 10.00 & 10.00 & 10.00 & 10.00 			& 10.00 & 10.00 & 10.00 & 10.00
		\\
		$i=j = $ 4-7   & 9.73  & 9.74  & 9.75  & 9.77  			& 9.70  & 9.73  & 9.75  & 9.76 
		\\ 
		$i=j = $ 8     & 9.70  & 9.71  & 9.72  & 9.73  			& 9.65  & 9.68  & 9.71  & 9.72 
		\\
		$i=j = $ 9-12  & 8.85  & 8.89  & 8.93  & 8.92  			& 9.21  & 9.26  & 9.31  & 9.32 
		\\
		$i=j = $ 13-14 & 8.84  & 8.86  & 8.90  & 8.90  			& 9.13  & 9.19  & 9.24  & 9.25 
		\\
		$i=j = $ 15    & 8.92  & 8.93  & 8.96  & 8.96  			& 9.23  & 9.27  & 9.31  & 9.32 
		\\
		$i=j = $ 16-19 & 8.36  & 8.38  & 8.40  & 8.40  			& 9.07  & 9.12  & 9.17  & 9.17 
		\\
		$i=j = $ 20-21 & 8.37  & 8.39  & 8.40  & 8.40  			& 9.04  & 9.09  & 9.15  & 9.15 
		\\
		$i=j = $ 22-23 & 8.33  & 8.34  & 8.36  & 8.36  			& 9.00  & 9.06  & 9.11  & 9.11 
		\\
		$i=j = $ 24    & 8.50  & 8.50  & 8.52  & 8.52  			& 9.12  & 9.17  & 9.22  & 9.22 
		\\
		$i=j = $ 0     & 9.28  & 9.29  & 9.30  & 9.30  			& 9.50  & 9.53  & 9.56  & 9.56 
		\\
		$i,j = $ 0,8   & 0.16  & 0.16  & 0.15  & 0.15  			& 0.19  & 0.17  & 0.16  & 0.15 
		\\
		$i,j = $ 0,15  & 0.48  & 0.48  & 0.47  & 0.47  			& 0.31  & 0.29  & 0.28  & 0.28 
		\\
		$i,j = $ 0,24  & 0.52  & 0.52  & 0.52  & 0.52  			& 0.25  & 0.24  & 0.23  & 0.23 
		\\
		$i,j = $ 8,15  & 0.11  & 0.10  & 0.10  & 0.09  			& 0.12  & 0.11  & 0.10  & 0.10 
		\\
		$i,j = $ 8,24  & 0.08  & 0.08  & 0.08  & 0.07  			& 0.09  & 0.09  & 0.08  & 0.08 
		\\
		$i,j = $ 15,24 & 0.24  & 0.24  & 0.23  & 0.23  			& 0.16  & 0.15  & 0.14  & 0.14 
		\\
		\hline
		$G_{11}$ 	   & 12.46 & 12.46 & 12.46 & 12.47 			& 11.12 & 11.05 & 10.98 & 10.99 
\\	\hline
	\end{tabular}
\end{table}
\FloatBarrier

\begin{table}[ht]
	\caption{
		\small
		Flavor-dependent  coupling constants $G_{ff}^{ps}$ and 
$G_{ff}^{s}$
		for normalized values of $G_{ij}^n$ by using the values from Table  (\ref{table:Gps}).
	} 
	\label{table:Gffps}
	\centering
	\begin{tabular}{ | l  | c c c c || c | c c c c |} \hline \centering
		Set 							& 1     & 2     & 3     & 4 	& 							& 1     & 2 	& 3     & 4 	
		\\ \hline
		$G_{uu}^{ps}/2$ (GeV$^{-2}$) 	& 10.00 & 10.00 & 10.00 & 10.00 & $G_{uu}^s/2$ (GeV$^{-2}$) & 10.00 & 10.00 & 10.00 & 10.00 
		\\
		$G_{ss}^{ps}/2$ (GeV$^{-2}$) 	& 9.55  & 9.57  & 9.58  & 9.60 	& $G_{ss}^s/2$ (GeV$^{-2}$) & 9.48  & 9.52  & 9.56  & 9.58 
		\\ 
		$G_{cc}^{ps}/2$ (GeV$^{-2}$) 	& 8.61  & 8.63  & 8.66  & 8.66 	& $G_{cc}^s/2$ (GeV$^{-2}$) & 9.02  & 9.08  & 9.14  & 9.14 
		\\
		$G_{bb}^{ps}/2$ (GeV$^{-2}$) 	& 8.24  & 8.24  & 8.25  & 8.25 	& $G_{bb}^s/2$ (GeV$^{-2}$) & 9.00  & 9.05  & 9.11  & 9.10 
		\\ \hline
	\end{tabular}
\end{table}
\FloatBarrier

\subsection{ Pseudoscalar  and scalar mesons spectrum and observables }

Numerical results
for pseudoscalar mesons masses, meson-quark coupling constants,
charged mesons weak decay constants and scalar quark-antiquark chiral condensates,
 are shown in Table (\ref{table:PSMesMas-M1}) for the 
different SETs of parameters.
For each of the SET of parameter three calculations of 
meson observables  are presented:
\begin{itemize}
\item
with $G_0$: outside parenthesis and brackets, 
\item
with $G_{ij}$: (inside parenthesis),
\item   
the limit of zero HQ condensates $\{ \bar{S}_c = \bar{S}_b \to 0\}$:
 $\{ \mbox{inside curly brackets}\}$.
\end{itemize}
 
Values in bold characters were those values obtained from the 
fitting procedure for the corresponding SET of parameters.
For example, parameters of SET 1 were found from a calculation with standard NJL with $G_0$
and the corresponding results in the first column of results are presented
with bold characters.
 By considering the same SET 1 of parameters but gap equations and BSE
for $G_{ij}$ results are presented inside parenthesis.
  The limit of zero HQ condensates was considered only for two SETs of parameters
(3 and 4)
just to obtain an estimation of their effects.
 The pseudoscalar mesons masses are in reasonable good agreement with Exp. values,
within  around $5\%$.
The mass of the pion does not change for each fitting because it was taken to normalize the coupling constants with $G_{11}$.
The flavor dependent coupling constants 
yield {\color{red}
- in general - 
slight  
increase (decrease) 
of the light (heavy) } meson masses not only  because  the corresponding 
shift in the quark effective masses but also due to their effects in the corresponding
BSE.
The corresponding 
meson-quark coupling constants   decrease with the use of the 
flavor dependent coupling constants. 
 SET  1 (3) presents the smaller (largest) values for the quark-meson coupling constants.

For the SETs 3 and 4 the effect of switching off the HQ condensates
are presented inside curly brackets.
Their effects  appear basically for the heavy meson masses as expected.
In the limit of zero HQ condensates the HQ constituent quark masses
equal their current masses and the heavy mesons masses  shift
nearly proportionally, according  to the basis of the constituent  quark model.
These results provide an indication of the relative role of the heavy quark condensates
for the meson masses and weak decay constants, 
that   many times have been argued to be zero as discussed above.
As an outcome, it can be seen 
that if the heavy quark condensates calculated in the NJL model 
could have smaller values (but non zero) 
the agreement with mesons masses would be better.
Therefore,  the NJL gap equations, in the mean field level,
may be seen  in a first analysis as
overestimating
the heavy quark-antiquark scalar condensates.
Nonetheless, we did not find a single value for $\left<\bar{c} c\right>$ and a single value for $\left<\bar{b} b\right>$ that 
reproduce uniformly the values of all the (pseudoscalar and scalar)  meson masses.

The 
 weak decay constants   of charged pseudoscalar mesons ($f_{ps}$)
are also presented.
NJL-type models usually do not reproduce these observables reasonably well
\cite{PNJL-D,HM-NJL-guo-etal} 
and this problem remains for the model as defined in this work
even with the variation of all the parameters considered.
It has been argued that considerably smaller coupling constants $G_0$
can reach values close to experimental ones for the heavy $B, B_s$ \cite{npb-ebert-etal-1995}
however we did not perform an arbitrary and independent variation of coupling constants.
Besides  usual  values for 
$f_\pi$ and $f_K$,
the decay constants $f_{D}$ and $f_{B_s}$ are reasonably
good  considering the usual results for the NJL model, all the other values
are in disagreement with experimental values.
The fitting procedure was also modified to make the $f_{ps}$ 
fitting observables.
However,  
it was not possible to find an unique set of  $M_f$ that 
provide reasonably good values for $f_{ps}$.
This shows that further physical input is needed, most probably 
by means of non-relativistic heavy quark approach or other interactions
\cite{narison-Fps,narison-Fps2,HM-decay-lucha-etal,mixing-etas,extrapolation-Fps}.

The quantities $\bar{S}_f$, representing the quark-antiquark 
scalar chiral condensates, were defined as:
\begin{eqnarray}
\bar{S}_f = - \left< \bar{q}_f q_f \right>^{\frac{1}{3}},
\end{eqnarray}
They  are presented
and compared to lattice calculations for the case of 
light $u,d$ and $s$ quarks and with improvements over 
QCD sum rule  for the heavy ones
\cite{anton+ribeiro}.
Results show a nearly monotonic increase  of the chiral condensates
with the quark mass that is  not necessarily found in other calculations.
The flavor dependent coupling constants may in some cases
 lead to a very  small decrease of the 
quark condensates, although these shifts 
are not as large as the shifts in the quark effective masses.
The SETs defined with flavor dependent 
cutoff in general  lead to larger values for the  
chiral condensates.

\begin{table}[ht]
		\caption{
 Pseudoscalar mesons masses and observables for the sets of parameters 
			 shown in Table (\ref{tab:parameters-gap})
			and the corresponding 
			coupling constants of Table (\ref{table:Gffps}).
The experimental or expected value (lattice QCD) are also shown (Exp.).
$^\dagger$The  pion mass displayed is an averaged value of the neutral and charged masses.
$\bar{S}_f = (-\left< \bar{q}_fq_f \right>)^{1/3}$. 
			} 
		\centering 
		\begin{tabular}{l c c c c c} 
			\hline\hline 
			Set				  & 1	   					& 2	  						& 3	 						& 4							& Exp. \\
			\hline \hline
			\\
			$M_{\pi}$ (MeV)   & {\bf 143}(143)  		& 142({\bf 142})			& {\bf 147}(147)			& 140({\bf 140})			& 137$^\dagger$ \\
							  & 						& 							& \{ 147(147) \} 			& \{ 140(140) \}  			& 
\\
			\hline
			\\
			$M_{K}$ (MeV) 	  & {\bf 484}(486)		   	& 493({\bf 494})	  		& {\bf 512}(513)	    	& 492({\bf 493})	    	& 495  \\
							  & 						& 							& \{ 512(513) \} 			& \{ 492(493) \} 			& 
\\
			\hline
			\\
			$M_{D}$ (MeV) 	  & {\bf 1846}(1841)	 	& 1867({\bf 1863})		 	& {\bf 1868}(1874)			& 1865({\bf 1856})		   	& 1870 \\
							  & 						& 							& \{ 1310(1381) \} 			& \{ 1201(1269) \} 			& 
\\
			\hline
			\\
			$M_{D_s}$ (MeV)   & {\bf 1986}(1966)	 	& 2009({\bf 1989})			& {\bf 2018}(2007)			& 2003({\bf 1978})		   	& 1968 \\
							  & 						& 							& \{ 1469(1520) \} 			& \{ 1351(1401) \} 			& 
\\
			\hline
			\\
            $M_B$ (MeV)		  & {\bf 5245}(5233)	 	& 5272({\bf 5260})			& {\bf 5279}(5280)			& 5268({\bf 5251})		  	&  5280 \\
							  & 						& 							& \{ 4740(4834) \} 			& \{ 4628(4722) \} 			& 
\\
            \hline
            \\
            $M_{B_s}$ (MeV)	  & {\bf 5379}(5353)	 	& 5407({\bf 5380})			& {\bf 5421}(5405)		  	& 5399({\bf 5366})	    	&  5367 \\
							  & 						& 							& \{ 4882(4959) \} 			& \{ 4759(4836) \} 			& 
\\
            \hline
            \\
            $M_{B_c}$ (MeV)	  & {\bf 6511}(6441)	 	& 6541({\bf 6467})			& {\bf 6539}(6482)			& 6513({\bf 6426})	    	& 6275 \\
							  & 						& 							& \{ 5491(5595) \} 			& \{ 5276(5379) \} 			& 
\\
			\hline \hline
			\\
			$g_{\pi qq}$ 	  & {\bf 3.35}(3.35) 		& 3.56({\bf 3.56})			& {\bf 3.78}(3.78)			& 3.78({\bf 3.78})			&  -  \\
 			$g_{Kqq}$ 		  & {\bf 3.59}(3.46)		& 3.80({\bf 3.66})			& {\bf 4.02}(3.89)			& 4.01({\bf 3.88})			&  -  \\
 			$g_{Dqq}$ 		  & {\bf 4.83}(4.20)		& 5.07({\bf 4.43})			& {\bf 5.31}(4.67)			& 5.29({\bf 4.64})			& -   \\
			$g_{D_sqq}$ 	  & {\bf 5.13}(4.44)		& 5.36({\bf 4.64})			& {\bf 5.59}(4.87)			& 5.56({\bf 4.84})			& -   \\
 			$g_{Bqq}$ 		  & {\bf 7.31}(6.03)		& 7.63({\bf 6.31})			& {\bf 7.95}(6.59)			& 7.92({\bf 6.56})			&  -  \\
 			$g_{B_sqq}$ 	  & {\bf 7.67}(6.33)		& 7.97({\bf 6.57})			& {\bf 8.27}(6.85)			& 8.22({\bf 6.80})			& -   \\
			$g_{B_cqq}$ 	  & {\bf 8.69}(7.14)		& 8.98({\bf 7.40})			& {\bf 9.27}(7.66)			& 9.25({\bf 7.62})			&  -  \\
			\hline
			\\
 			$f_{\pi}$ (MeV)	  & {\bf 116}(116)			& 119({\bf 119})			& {\bf 122}(122)			& 122({\bf 122})			& 92   \\
							  & 						& 							& \{ 122(122) \} 			& \{ 122(122) \} 			& 
\\
 			$f_{K}$ (MeV)	  & {\bf 119}(119)			& 121({\bf 121})			& {\bf 123}(123)			& 123({\bf 123})			& 110  \\
							  & 						& 							& \{ 123(123) \} 			& \{ 123(123) \} 			& 
\\
 			$f_{D}$ (MeV)	  & {\bf 144}(142)			& 143({\bf 142})			& {\bf 143}(142)			& 142({\bf 140})			& 150  \\
							  & 						& 							& \{ 134(134) \} 			& \{ 131(131) \} 			& 
\\
 			$f_{D_s}$ (MeV)	  & {\bf 122}(123)			& 123({\bf 124})			& {\bf 124}(125)			& 124({\bf 125})			& 177  \\
							  & 						& 							& \{ 120(121) \} 			& \{ 119(120) \} 			& 
\\
 			$f_B$ (MeV)		  & {\bf 205}(203)			& 200({\bf 198})			& {\bf 197}(195)			& 195({\bf 193})			& 134  \\
							  & 						& 							& \{ 189(189) \} 			& \{ 186(186) \} 			& 
\\
 			$f_{B_s}$ (MeV)	  & {\bf 161}(164)			& 160({\bf 163})			& {\bf 159}(162)			& 159({\bf 162})			& 162  \\
							  & 						& 							& \{ 153(157) \} 			& \{ 153(156) \} 			& 
\\
 			$f_{B_c}$ (MeV)	  & {\bf 78.8}(80.5)		& 80.3({\bf 82.1})			& {\bf 82.3}(83.9)			& 82.4({\bf 84.4})			& 302  \\
							  & 						& 							& \{ 98.4(98.4) \} 			& \{ 103(103) \} 			&   
\\
			\hline
			\\
 			$\bar{S}_u$ (MeV) & {\bf 339}(339)			& 348({\bf 348})			& {\bf 358}(358)			& 358({\bf 358})			& 283 \cite{flag,lightcondensate}  \\
							  & 						& 							& \{ 358(358) \}			& \{ 358(358) \}			& 
\\
			$\bar{S}_s$ (MeV) & {\bf 356}(354)			& 364({\bf 362})			& {\bf 373}(371)			& 372({\bf 371})			& 290 \cite{flag,lightcondensate} \\
							  & 						& 							& \{ 373(371) \}			& \{ 372(371) \}			& 
\\
			$\bar{S}_c$ (MeV) & {\bf 375}(374)			& 382({\bf 382})			& {\bf 375}(374)			& 396({\bf 396})			& 196/180 \cite{anton+ribeiro} \\
							  & 						& 							& \{ 0.0(0.0) \}			& \{ 0.0(0.0) \}			& 
\\
			$\bar{S}_b$ (MeV) & {\bf 377}(377)			& 384({\bf 384})			& {\bf 377}(377)			& 399({\bf 399})		 	& 219/213 \cite{anton+ribeiro}    \\
							  & 						& 							& \{ 0.0(0.0) \}			& \{ 0.0(0.0) \}			&  
\\
			\hline \hline
		\end{tabular}
		\label{table:PSMesMas-M1} 
	\end{table}
\FloatBarrier

From the shifts in the values of the quark-meson coupling constants shown in the previous Table,
it is possible to read the probability of a meson with valence quark/antiquark $q_1,q_2$
develop different flavor sea quarks (antiquarks) since 
$g_{psqq} = Z^2_{M}$ that is the pseudoscalar meson renormalization constant.
 The original valence quark-antiquark structure is encoded
in the value calculated for the coupling constant of reference $G_0$
and the probability of appearing different sea-quark-antiquark components
 was taken as the percentage deviation of $Z_M$.
Results are shown in Table (\ref{table:probabilities}).
The pion was kept unmodified due to the 
(re)normalization procedure.
There is a  monotonic trend to 
 larger  new contributions  in the heavier mesons.
It  may be  nearly 
in  qualitative agreement with similar conclusions from the analysis of 
energy distribution amplitudes \cite{nonvalence}.
 
\begin{table}[ht]
		\caption{
Probabilities of a meson with valence quark-antiquark structure to  develop other types of 
sea quark/antiquark components from the results for $g_{psqq}$ presented in Table 
(\ref{table:PSMesMas-M1}).
			} 
		\centering  
		\begin{tabular}{l c c c c }  
			\hline\hline  
			Set				  & 1	   					& 2	  
		& 3	 						& 4							\\
			\hline \hline
			\\
			$Pr(\pi)$ 	  &  -  & -   	& -  & -  \\
 			$Pr(K)$		  & 1.8	 $\%$ & 1.9	 $\%$ &  1.6  $\%$	&  1.6  $\%$ 	   \\
 			$Pr(D)$		  &  7.2  $\%$	&  6.5 	 $\%$ &  6.2  $\%$	& 6.3  $\%$    \\
			$Pr(D_s)$	  &    7.0 $\%$	& 7.0  $\%$ 	&  6.7   $\%$  & 6.7  $\%$	  \\
 			$Pr(B)$		  & 9.2  $\%$	& 9.1 	 $\%$ &  8.9  $\%$  & 9.0   $\%$    \\
 			$Pr(B_s)$  &  9,2 $\%$	& 9.2 	 $\%$ &  9.0  $\%$  & 9.1 $\%$   \\
			$Pr(B_c)$ 	  & 9.4  $\%$	&  9.2  $\%$	& 9.1  $\%$ 	& 9.2  $\%$    \\
			\hline
\hline
		\end{tabular}
		\label{table:probabilities}  
	\end{table}
\FloatBarrier

In spite of the problems usually found
to  identify many of the known  scalar mesons to 
quark-antiquark structures, at least as main components, 
the  masses of the scalar quark-antiquark states of scalar mesons
 were calculated according to structures shown in Table (\ref{table:comp}) and 
results are shown 
in Table (\ref{tab:scalars}) - except for the states built up with $S_i$ ($i=0,8,15,24$, called here as
{\it diagonal})
for the SETs of parameters defined above.
Some 
experimental \cite{PDG} or expected  \cite{flag} values (Exp.) 
are also shown. 
For the masses of the scalar states, an infrared cutoff 
to avoid the threshold of the unphysical decay on a pair of quark-antiquark was considered. 
Effectively this IR cutoff prevents long range modes to contribute, being therefore
associated to confinement.
Seen otherwise, 
this procedure is nearly equivalent to increasing quark current masses 
such as to produce large quark mass
confinement.
Their values were not used to fit the mesons masses, but rather were extracted
from the BSE as the minimum values for which the BSE provides a result safe from
the unphysical decay.
This IR cutoff can be associated to the longest wavelength mode 
that would contribute for the meson structure, and therefore it may be argued to 
be inversely proportional to a sort of  {\it  meson radius}
that could be compared to predictions for the averaged quadratic radia \cite{charge-radia}.
The following values of IR cutoffs were considered, 
with a corresponding naif 
value of a {\it meson radius} ($\Lambda_{IR} \sim 1/R_{mes}$) in parentheses
for light $(l)$ and heavy $(h)$ quark structure:
\begin{eqnarray}
\Lambda_{IR}^{ll} = 95 \mbox{MeV}, \;\;  ( 1.04 \mbox{fm}) , \;\;\;\;\;\;\;\;
\Lambda_{IR}^{lh} = 160 \mbox{MeV},  \;\;  ( 0.61 \mbox{fm}) ,  \;\;\;\;\;\;\;\;
\Lambda_{IR}^{hh} = 290 \mbox{MeV} \;\;  ( 0.34 \mbox{fm}) .
\end{eqnarray}

Results for the  light  neutral scalars $a_0$ and $K_0^*$
present the usual inversion of hierarchy of NJL-type models \cite{scalars-light,pelaez-status,NJL-scalars}.
This inversion also happens for the $D_0^*$ and $D_{s0}^*$
but not for $B_0^*, B_{s0}^*$ and $B_{c0}^*$.
For all these heavy mesons  the discrepancy of their masses with respect to 
the Exp. values  is nearly  
around $5\%$
(or in some cases $10\%$).
The calculations with $G_{ij}$  lead to masses slightly smaller 
or larger than
calculations with $G_0$.
SETs fixed by considering $G_{ij}$ however, usually, present larger meson 
masses than those SETs with parameters
fixed with $G_0$.
By switching off the HQ condensates the heavy meson masses also decrease
nearly proportionally to the corresponding decrease of the constituent 
quark effective masses.
The SETs with flavor-dependent cutoffs also  present larger values of masses.

\begin{table}[ht]
		\caption{
			\small 
Masses of scalar (quark-antiquark) states or 
 mesons   for the same sets used for Table
(\ref{table:PSMesMas-M1}).
Exp:
experimental values from  PDG 
and, for the heaviest ones, results from lattice QCD \cite{flag}.			} 
		\centering 
		\begin{tabular}{l c c c c c} 
			\hline\hline 
			Set				     & 1						& 2	   						& 3	  						& 4	 						& Exp. \\
			\hline \hline
			\\
			$M_{a_0}$ (MeV)      & {\bf 540}(540)			& 628({\bf 628})			& {\bf 718}(718)			& 714({\bf 714})			& 980 \cite{PDG} \\
								 & 							& 							& \{ 718(718) \} 			& \{ 714(714) \} 			& 
\\
			\hline
			\\
			$M_{K_0^*}$ (MeV)    & {\bf 867}(843)			& 936({\bf 910})			& {\bf 1017}(991)			& 998({\bf 975})			& 700-900 \cite{PDG} \\
								 & 							& 							& \{ 1017(991) \} 			& \{ 998(975) \} 			& 
\\
			\hline  
			\\
			$M_{D_0^*}$ (MeV)    & {\bf 2148}(2121)			& 2213({\bf 2185})			& {\bf 2267}(2243)			& 2252({\bf 2220})			& 2343 \cite{PDG} \\
								 & 							& 							& \{ 1694(1722) \} 			& \{ 1564(1593) \} 			& 
\\			\hline
			\\ 
			$M_{D_{s0}^*}$ (MeV) & {\bf 2404}(2335)			& 2464({\bf 2395})			& {\bf 2519}(2456)			& 2491({\bf 2421})			& 2318 \cite{PDG} \\
								 & 							& 							& \{ 1962(1948) \} 			& \{ 1823(1814) \} 		 	& 
\\
			\hline
			\\ 
            $M_{B_0^*}$ (MeV)    & {\bf 5568}(5534)			& 5634({\bf 5599})			& {\bf 5687}(5656)			& 5668({\bf 5629})			& 5730 \cite{flag} \\
								 & 							& 							& \{ 5148(5166) \} 			& \{ 5028(5047) \} 			& 
\\
			\hline
			\\ 
            $M_{B_{s0}^*}$ (MeV) & {\bf 5798}(5721)			& 5861({\bf 5783})			& {\bf 5915}(5844)		 	& 5883({\bf 5806})			& 5776 \cite{flag} \\
								 & 							& 							& \{ 5378(5356) \} 			& \{ 5245(5226) \} 			& 
\\
			\hline
			\\ 
            $M_{B_{c0}^*}$ (MeV) & {\bf 7040}(6936)			& 7102({\bf 6999})			& {\bf 7135}(7045)			& 7105({\bf 6997})			& 6711 \cite{flag} \\
								 & 							& 							& \{ 6058(6068) \} 			& \{ 5821(5835) \} 			& 
\\
			\hline \hline
		\end{tabular}
		\label{tab:scalars} 
	\end{table}
\FloatBarrier

The masses of the {\it diagonal}
pseudoscalar and scalar states,
i.e. built up with the diagonal
generators as indicated
in Eq. (\ref{mesons-0-8-etc}),
are presented in Table 
(\ref{table:PSMesMas-M3}).
For these  scalar mesons BSE,  the following IR cutoffs were used:
\begin{eqnarray}
\Lambda_{IR}^{ll} = 170 \mbox{MeV}, \;\;\;\;\;\;
\Lambda_{IR}^{lh} = 470 \mbox{MeV},  \;\;\;\;\;\;
\Lambda_{IR}^{hh} = 470 \mbox{MeV}.
\end{eqnarray}
These IR cutoffs, larger than the 
cutoffs used for the other scalar mesons, were also fixed as the minimum values to provide stable solutions for the scalar mesons BSE
avoiding their unphysical decays.

The flavor-dependent coupling constants can lead to either a decrease ot increase of the 
 mesons masses.
 Although the mass of the $\eta$ 
 is surprisingly good, the $\eta'$ mass is not correct.
 This may seem a  simple corroboration of the need of the determinantal 't Hooft interaction 
 and / or 
the calculation with the mixing interactions $G_{i\neq j}$
as presented in \cite{PRD-2021,JPG-2022} for the light mesons.
The masses of the heavy pseudoscalar mesons $\eta_c, \eta_b$,
and also of the heavy scalar mesons $\chi_{c0},\chi_{b0}$,  are 
larger than expected with a discrepancy inside nearly   $10 \%$.
In
 the limit in which the heavy quark condensate is switched off, their masses 
become too small. 
This goes along with the conclusion  that possibly  the 
heavy quark condensates are overestimated
in the NJL model at the mean field level.
For the fits used in this work,  the masses of the 
controversial light mesons $f_0(500)$ and $f_0(980)$ can  be - not simultaneously - close to their 
experimental value  depending on the SET of parameters.
The heavy scalars masses ($\chi_{c0}$ and $\chi_{b0}$) are also too large
although within around $7\%$ when compared to Exp. values.
As a consequence, some of the usual problems to describe light scalar mesons by considering 
only the quark-antiquark states  in the NJL model persists.
These scalar neutral mesons are usually argued  
to contain  larger gluonic components 
or other combinations of neutral flavor tetraquark currents or even 
flavor  molecules.

\begin{table}[ht]
	\caption{
		\small Masses of pseudoscalar and 
		scalar mesons 
		built up with the diagonal generators
		according to Table (\ref{table:comp})  without any mixing effect or coupling, 
		neither in the polarization tensor,  Eqs.  (\ref{table:Gij}), nor in the 
 $G_{ij}^n$, Eqs. (\ref{Guu}-\ref{Gbb}).
		Experimental values (Exp.) extracted from PDG \cite{PDG}.
	} 
	\centering 
	\begin{tabular}{c | c c c c | c } 
		\hline\hline 
		Set 								& 1 					&  2 					& 3 					& 4 					& Exp.  \cite{PDG}
		\\
		\hline \hline
		$M_{PS,0}$ ($M_{\eta'}$) (MeV)		& {\bf 697}(723) 		& 738({\bf 766})  		& {\bf 782}(813)   		& 777({\bf 808})  		& 958 \\
											&						&						& \{ 764(797) \} 		& \{ 753(787) \} 		& 
		\\
		$M_{PS,8}$ ($M_{\eta}$) (MeV)      	& {\bf 501}(509)  		& 516({\bf 523})  		& {\bf 539}(546)    	& 522({\bf 529})    	& 548 \\
											&						&						& \{ 539(546) \} 		& \{ 522(529) \} 		&  
		\\
		$M_{PS,15}$ ($M_{\eta_c}$) (MeV)    & {\bf 3321}(3221)  	& 3360({\bf 3257}) 		& {\bf 3370}(3279) 		& 3353({\bf 3236}) 		& 2984 \\
											&						&						& \{ 2366(2400) \} 		& \{ 2178(2207) \} 		&  
		\\
		$M_{PS,24}$ ($M_{\eta_b}$) (MeV)    & {\bf 9964}(9843) 		& 10003({\bf 9874})		& {\bf 10010}(9896)		& 9973({\bf 9826})		& 9399 \\
											&						&						& \{ 8940(9012) \} 		& \{ 8704(8777) \} 		& 
\\
\hline		
		$M_{S,0}$ ($M_{f_0(980)}$) (MeV) 	& {\bf 782}(797)		& 852({\bf 865}) 		& {\bf 930}(941)		& 920({\bf 932})      	& 990 \\
											&						&						& \{ 888(906) \} 		& \{ 854(875) \} 		& 
		\\
		$M_{S,8}$ ($M_{f_0(500)}$) (MeV)  	& {\bf 598}(581)		& 682({\bf 662})		& {\bf 777}(758)		& 752({\bf 734}) 		& 500 \\
											&						&						& \{ 777(758) \} 		& \{ 752(734) \} 		& 
		\\ 
		$M_{S,15}$ ($M_{\chi_{c0}}$) (MeV)	& {\bf 3583}(3496)		& 3638({\bf 3552}) 		& {\bf 3670}(3595)		& 3636({\bf 3543}) 		& 3415 \\
											&						&						& \{ 2478(2513) \} 		& \{ 2223(2255) \} 		& 
		\\
		$M_{S,24}$ ($M_{\chi_{b0}}$) (MeV)	& {\bf 10412}(10306)	& 10476({\bf 10369})	& {\bf 10508}(10414)	& 10473({\bf 10360})	& 9859 \\
											&						&						& \{ 9437(9440) \} 		& \{ 9202(9205) \} 		& 
		\\
		\hline \hline
	\end{tabular}
	\label{table:PSMesMas-M3} 
\end{table}
\FloatBarrier

\section{ Final remarks}

The NJL  model
provides a quite simple framework to describe
 mesons   as quark-antiquark states
being, in the present work,  extended for flavor  U(5)  with 
flavor dependent coupling constants   and cutoffs.
The resulting (normalized) coupling constants, $G_{ij}$, present the 
same  expected trend  of  early analysis from light flavor NJL model
and they naturally lead to a distinction between the behavior of  light and heavy quark components
in agreement with previous expectations
\cite{npb-ebert-etal-1995}.
Mixing interactions, $G_{i \neq j}$ and $G_{f_1 \neq f_2}$,
 have not  been considered fully. 
Being considerably smaller than the  diagonal interactions, $G_{ii}$ and $G_{ff}$,
their normalizations are not easily defined.
  A more extensive and complete analysis of the effects of the mixing interactions 
depends on the ambiguity of their normalizations and this issue was outside the scope of the present
 work.
Seven of these pseudoscalar meson masses were used as fitting observables.
Surprisingly 
 all the pseudoscalar and, almost all of the, scalar meson masses,
for degenerate up and down quark effective masses, were reasonably well reproduced  
with overall accuracy of predictions around $5\%-10\%$.
The only exceptions were the masses of the light scalars $K_0^*(800)$ and $a_0(980)$ 
that  present the same inverted hierarchy as the standard NJL model, one of
the mesons $f_0(500)$ and $f_0(980)$ that are usually expected to not be 
exclusively quark-antiquark states and the $\eta'$ meson usually expected to rely on the 
effective interaction that accounts for the axial anomaly 
or the mixing interactions $G_{i \neq j}$.
 Other observables, however,   still need further improvements of the model,
eg. pseudoscalar meson weak decay constants (mainly the heavy ones)
 and seemingly the
(heavy) quark-antiquark condensates.
The flavor-dependent coupling constants  induced by polarization of the vacuum 
 introduce 
quantum mixings due to the different representation of the flavor group in which 
quarks and mesons are defined
although
it is not evident how to identify unambiguously the individual  
contribution of each species of these corresponding sea quarks/antiquarks (u,s,c and b) 
in a particular one  of these coupling constants $G_{ff}$.

The pseudoscalar mesons built up with the diagonal flavor generators,
$\lambda_i$ for $i=0,8,15,24$ are usually labeled as $\eta's$ ($\eta, \eta',\eta_c,\eta_b$).
Although the mass of the light $\eta$ can be reproduced by these coupling constants
without a mixing interaction with the $\eta'$, 
the mass of the $\eta'$ is not fully obtained in this approach - the best estimates yield
$m_{\eta'} \simeq 800$MeV.
The masses of the  heavy diagonal pseudoscalar   ($\eta_c,\eta_b$) 
agree  with the experimental values within  around $7\%$ - or less.
In general, the obtained values for the 
heavy $\eta$ masses suggest  that the NJL gap equations -  as discussed in this 
usual mean field treatment -  overestimate the heavy quark condensates..
This conclusion is somehow also taken from several of the other pseudoscalar meson masses.
We did not find an unique set of heavy quark effective masses,
or conversely quark  chiral condensates,  that produce the best 
agreement of masses with experimental values.
Although the structure of the 
 light scalars, within the flavor U(3) description, have been in a longstanding controversial 
debate,  one can assume that the quark-antiquark states can be considered
part of these mesons.
An effective  IR cutoff was introduced for the BSE of the scalar  that
prevents long range modes to contribute, being therefore
associated to confinement.
This procedure is nearly equivalent to increasing quark current masses 
such as to produce large quark mass
confinement. 
Although two or three IR cutoffs were introduced,
they were not chosen to fit the scalar meson masses but rather
to avoid the unphysical decay in a pair of quark-antiquark.
It turns out however, that the masses of the scalars calculated with this IR cutoff
are reasonably good.
In spite of the difficulty in describing the masses of 
the light $a_0(980)$, $K_0^*(800)$ and  $f_0(500)$ or $f_0(980)$ (that 
are not fitted by a same set of parameters),
the predictions for the other scalar meson masses agree within around
$10\%$. 
Indeed, the structures of  these mesons  are highly controversial 
with strong evidences for the need of non-quark-antiquark states \cite{pelaez-status}.

The effects of the flavor dependent $G_{ij}$ were analyzed by switching them on
or off,  $G_0 \leftrightarrow G_{ij}$, and this leads to a small decreases of 
quark effective masses and   mesons masses when using $G_{ij}$.
This happens in both cases: when $G_0$ is considered for the fitting procedure and 
when $G_{ij}$ is considered for the fitting.
The usual difficulty to reproduce the  pseudoscalar meson weak decay constants for the 
NJL models
appears to be worse for the heavy mesons.
The only heavy meson weak decay constants that  are reasonably well described
are the $D$ and $B_s$,  the numerical values of all the other heavy meson
decay constants are far from the 
experimental/expected value, with an incorrect hierarchy of values.
The quite good agreement of meson masses
 show that heavy quark symmetries
  imposed mainly in the heavy quark propagator and interactions
may not 
be relevant for their predictions within the NJL model,
 although other observables such as the 
weak decay constants should receive important modifications.
The NJL-model is  a non renormalizable effective model
and  there is a need of an energy scale 
such as an UV cutoff.
 The non-covariant three-dimensional
cutoff scheme was adopted
being that the values for heavy and light quarks
were found to be  nearly the same 
with a quite low value when comparing
to most of the meson masses  and compared to the heavy quark masses.
This choice for the regularization scheme 
seems to be relevance for 
the quite good agreement of fitted and  predicted meson masses
basically without the need of flavor dependent UV cutoffs.
The present regularization scheme makes the interpretation of 
the UV cutoff for the light and heavy quarks   uniform 
and it favors an unique cutoff framework.
It would be very interesting to understand further   this issue for different regularization schemes. In other words,  can the flavor dependence of the coupling constants eliminate the need of 
flavor dependent cutoffs in covariant regularization schemes?
There are some other aspects that can  be investigated further. Firstly, 
a self-consistent calculation of 
effective masses and coupling constants must be done with the corresponding fitting of mesons masses and weak decay constants. In this respect, the mixing interactions can be expected to have non-trivial consequences for both the gap equations solutions,  for the {\it diagonal} mesons (both scalars and pseudoscalars as usually discussed in the literature) and for the 
weak decay constants.
The role of non-relativistic reduction of the NJL model would be of interest, for example, to verify its role on the pseudoscalar meson decay constants. 
After all, it looks it is missing some physical input for the description of the $f_{ps}$.

\section*{Acknowledgements}

F.L.B. is member of
INCT-FNA,  Proc. 464898/2014-5
and  he acknowledges partial support from 
 CNPq-312750/2021-8,
 CNPq-312072/2018-0 
and  
CNPq-421480/2018-1.
The authors thank short conversations
 with C.D. Roberts, B.El Bennich, T. Frederico, G.I. Krein,
 F.S. Serna  and A. Pineda.

\appendix

\section{ Appendix:  The quark propagator in the adjoint representation}
\label{app:quarkprop}

\renewcommand{\theequation}{A.\arabic{equation}}

The following parameterization of the 
quark masses matrix,  both current and constituent masses,
was considered for the calculation of traces in the flavor index:
\begin{eqnarray}
M_f  &=& \sum_{i=0,8,15,24} \lambda_i M_i,
\\
\label{mass-adjoint}
M_0 &=& \frac{1}{\sqrt{10}} ( 2 M_u + M_s + M_c + M_b),
\nonumber
\\
M_8 &=& \frac{1}{\sqrt{3}} ( M_u - M_s ),
\nonumber
\\
M_{15} &=& \frac{1}{2 \sqrt{6}} ( 2 M_u + M_s  +  3 M_c ),
\nonumber
\\
M_{24} &=& \frac{1}{ 2 \sqrt{10} } ( 2 M_u + M_s + M_c  - 4 M_b ).
\end{eqnarray}
 And the same parameterization was done  for the quark propagators ($i$) such that:
\begin{eqnarray}
S_f (k) = \sum_i \lambda_i  \cdot S_i (k) ,
\end{eqnarray}
where $S_f (k) = 1/(\slashed{k} - M_f)$
being that these sums in the index $_i$ are the same as those in Eqs. (\ref{mass-adjoint}) for each 
$i=0,8,15,24$.


\begin{thebibliography}{99}


 





 
\bibitem{yan-etal}
Tung-Mow Yan,Hai-Yang Cheng, Chi-Yee Cheung, and Guey-Lin Lin,
Y. C. Lin, Hoi-Lai Yu,
Heavy-quark symmetry and chiral dynamics,
Phys. Rev. D46, 1148 (1992)

\bibitem{nowak-etal}
M. A. Nowak, M. Rho, and I. Zahed,
Chiral effective action with heavy-quark symmetry,
Phys. Rev. D48, 4370 (1993)


\bibitem{HQ-review}
 A. Manohar and M. Wise, Heavy Quark Physics (Cambrige University Press, 2000).
 M. Neubert, Phys. Rept. 245, 259 (1994), hep-ph/9306320.

 


\bibitem{review-HQS}
M. Neubert, Heavy-quark symmetry,
Phys. Rept. 245, 259 (1994).



\bibitem{isgur-wise}
 N. Isgur and M. B. Wise, Phys. Lett. B232, 113 (1989);
B237, 527 (1990).

\bibitem{HQ-observ}
 M. A. Ivanov, Y. L. Kalinovsky, P. Maris, and C. D. Roberts, Phys. Lett. B416, 29 (1998), nucl-th/9704039.
 M. A. Ivanov, Y. L. Kalinovsky, and C. D. Roberts, Phys. Rev. D60, 034018 (1999), nucl-th/9812063.
 C. D. Roberts and S. M. Schmidt, Prog. Part. Nucl. Phys. 45, S1 (2000), nucl-th/0005064.


\bibitem{HQS-wise}
M.B. Wise,  
Combining Chiral and Heavy Quark Symmetries,
arXiv:hep-ph/9306277;
CALT-68-1860.

 

\bibitem{cheng-etal-S+A}
Hai-Yang Cheng, Fu-Sheng Yu,
Masses of scalar and axial-vector B mesons revisited,
Eur. Phys. J. C77, 668 (2017).




\bibitem{beneke}
M. Beneke, 
 Perturbative heavy quark-antiquark systems,
JEHP PRoceedings, Heavy Flavours 8, Southampton, UK, 1999.


\bibitem{HQS-ChDyn}
Tung-Mow Yan,Hai-Yang Cheng, Chi-Yee Cheung, and Guey-Lin Lin, Y. C. Lin,
Hoi-Lai Yu,
Heavy-quark symmetry and chiral dynamics,
 Phys.Rev. D 46, 1148 (1992).


\bibitem{review-CL}
R. Casalbuoni, {\it et al},
Phenomenology of Heavy Mesons Chiral Lagrangians,
Phys. Rept. 281, 145  (1997).
 
\bibitem{jiang-etal-2019}
S.-Z. Jiang, Y.R. Liu, Q.-H. Yang,
Chiral Lagrangians for mesons with a single heavy quark,
Phys. Rev. D 99, 074018 (2019).

\bibitem{yan-etal-1992}
T.-M. Yan {\it et al},
Heavy-quark symmetry and chiral dynamics,
Phys. Rev. D 46, 1148 (1992).

\bibitem{lattice}
S. Aoki et al, FLAG Review 2019, EPJ {\bf C 80}, 2 (2020).
O. Lakhina, E.S. Swanson, A canonical $D_s (2317)$?, Phys. Lett. B650,  159 (2007).

\bibitem{SDE}
For example:
Hui-Yu Xing, Zhen-Ni Xu, Zhu-Fang Cui, Craig D. Roberts, Chang Xu,
Heavy + heavy and heavy + light pseudoscalar to vector semileptonic transitions,
Eur. Phys. Journ. C 82,  889 (2022) .


\bibitem{strings}
E.L. Gubankova, A.Yu. Dubin,
Dynamical regimes of the QCD string with light and heavy quarks,
Phys. Lett. B 334, 180  (1994).


\bibitem{PRD-2022b}
T.H.Moreira, F.L.Braghin,
Magnetic field induced corrections to the NJL model coupling constant from vacuum polarization,
 Phys.  Rev. D 105,  114009
(2022).



\bibitem{NJL}
Y. Nambu, G. Jona-Lasinio, 
Dynamical Model of Elementary Particles Based on an Analogy with Superconductivity I,
Phys. Rev. {\bf 122},  345  (1961).


\bibitem{klevansky}
S. P. Klevansky,
The Nambu—Jona-Lasinio model of quantum chromodynamics,
 Rev. Mod. Phys. 64, 649 (1992).


\bibitem{vogl-weise}
U. Vogl, W. Weise, 
 The Nambu and Jona-Lasinio model: Its implications for Hadrons and Nuclei,
 Progr. in Part. and Nucl. Phys. {\bf 27}, 195   (1991).




\bibitem{hatsuda-etal}
T. Hatsuda and T. Kunihiro, Prog. Theor. Phys. 74 ( 1985 )
765; Prog. Theor. Phys. Suppl. 91, 284  (1987).



\bibitem{baryon-NJL1}
Yu-Ji Shi,
Hadronization of heavy diquark and light quark within NJL-Model,
arXiv:2005.08680v1 [hep-ph]

\bibitem{baryons-2}
W.M. Alberico, F. Giacosa, M. Nardi, and C. Ratti,
Baryonic masses based on the NJL model,
Eur. Phys. J. A 16, 221 (2003)

\bibitem{HM-NJL-guo-etal}
Xiao-Yu Guo, Xiao-Lin Chen, and Wei-Zhen Deng,
The heavy mesons in Nambu–Jona-Lasinio model
arXiv:1205.0355v1 [hep-ph]


\bibitem{npb-ebert-etal-1995}
D. Ebert, T. Feldmann, R. Friedrich, H. Reinhardt,
Effective meson lagrangian with chiral and heavy
quark symmetries from quark flavor dynamics,
Nuclear Physics B 434, 619  (1995).


\bibitem{gottfried-etal}
F.O. Gottfried and S.P. Klevansky,
Thermodynamics of open and hidden charmed mesons
within the NJL model,
 Physics Letters B 286, 221 (1992).


\bibitem{nam-prd}
Seung-il Nam, 
Extended nonlocal chiral-quark model for the D- and B-meson weak-decay constants,
Phys. Rev. D85, 034019 (2012)


\bibitem{bardeen-hill} 
W.A. Bardeen, C.T. Hill, 
Chiral dynamics and heavy quark symmetry in a solvable toy field-theoretic model,
Phys. Rev. D49, 409 (1994)


\bibitem{arriola}
A. L. Mota, E. Ruiz Arriola,
Relativistic NJL model with light and heavy quarks,
Eur. Phys. J. A 31,  711 (2007).

\bibitem{tt-btquarks}
W. A. Bardeen, C. T. Hill,  M. Lindner, 
Minimal dynamical symmetry breaking of the standard model,
Phys. Rev. D41, 1647  (1990).

\bibitem{QCDSR-nocond} 
D. Antonov, J.E.F.T. Ribeiro, Phys. Rev. D 81, 054027 (2010)
9. M.A. Shifman, A.I. Vainshtein, V.I. Zakharov, Nucl. Phys. B 147,
385 (1979).

\bibitem{NJL-t-b}
B. Durand, T. Zhang,
New solution for dynamical symmetry breaking with
top and bottom quark condensates,
arXiv:hep-ph/9408357v1.


\bibitem{anton+ribeiro}
D. Antonov, J. E. F. T. Ribeiro,
Quark condensate for various heavy flavors,
Eur.Phys.J. C72, 2179  (2012) 
arXiv:1209.0408 [hep-ph].
 


\bibitem{godfrey-isgur}
Stephen Godfrey and Nathan Isgur,
Mesons in a relativized quark model with chromodynamics,
Phys. Rev. D32, 189 (1985).
Stephen Godfrey and Richard Kokoski,
Properties of P-wave mesons with one heavy quark,
Phys. Rev. D43, 1679 (1991).

\bibitem{relativ-effects}
P. Colangelo, G. Nardulli, M. Pietroni,
Relativistic bound-state effects in heavy-meson physics,
Phys. Rev. D 43, 3002 (1991).




\bibitem{kohyama-etal}
H. Kohyama, D. Kimura, T. Inagaki,
Parameter fitting in three-flavor Nambu-Jona-Lasinio model with 
various regularizations,
Nucl. Phys. B906, 524 (2016).


 
\bibitem{SBK}
F. E. Serna, B. El-Bennich,  G. Krein,
Charmed mesons with a symmetry-preserving contact interaction,
Phys. Rev. D96, 014013 (2017).


\bibitem{mexicanos-2019}
L. X. Gutierrez-Guerrero, Adnan Bashir, Marco A. Bedolla, E. Santopinto,
Masses of Light and Heavy Mesons and Baryons: A Unified Picture,
Phys. Rev. D 100, 114032 (2019)
arXiv:1911.09213v1 [nucl-th].
 

\bibitem{bashir-etal}
Adnan Bashir et al,
Collective Perspective on Advances in Dyson–Schwinger Equation QCD,
Commun. Theor. Phys. 58,  79 (2012).

 
\bibitem{craig-etal}
Pei-Lin Yin, Zhu-Fang Cui, Craig D. Roberts, Jorge Segovia
Masses of positive- and negative-parity hadron ground-states,
including those with heavy quarks,
arXiv:2102.12568v1 [hep-ph]





\bibitem{fierz-transformation}
 C. Itzykson and J. B. Zuber, Quantum Field Theory, (McGraw-Hill, 1980).
M. Fierz, “Zur Fermischen Theorie des $\beta$-Zerfalls,” ,
Z. Physik 104, 553-565 (1937).
%

\bibitem{heavy-quark-pot}
J.L. Richardson,  
The heavy quark potential and the 
$\Upsilon$ and $J/\Psi$ system,
Phys. Lett. B 82, 272  (1979).


\bibitem{relat-heavy-pot}
P. Colangelo, G. Nardulli, M. Pietroni,
Relativistic bound-state effects in heavy-meson physics,
Phys. Rev. D43, 3002  (1991)



\bibitem{pattern-interaction}
Muyang Chen, Lei Chang,
A pattern for the flavor dependent quark-antiquark interaction,
Chinese Physics C Vol. 43,  114103  (2019).
M. Chen,
L. Chang, 
Y.-x. Liu,
Bc Meson Spectrum Via Dyson-Schwinger Equation and Bethe-Salpeter Equation
Approach,
arXiv:2001.00161v1 [hep-ph].


\bibitem{PRD-2021}
F.L. Braghin,
Flavor-dependent corrections for the U(3) NJL coupling constant,
Phys. Rev. D 103, 094028 (2021),
  arXiv:2008.00346v2 [hep-ph].



\bibitem{contact-lhc}
F. Bazzocchi, U. De Sanctis, M. Fabbrichesi, A. Tonero,
Quark contact interactions at the LHC,
Phys. Rev. D 85, 114001 (2012).
  


\bibitem{kaplan}
D.B. Kaplan, Lectures on effective field theory - ICTP-SAIFR, (2016).



\bibitem{JPG-2022} 
F.L.Braghin,
Strangeness content of the pion in the U(3) Nambu–Jona–Lasinio model,
 Journ. of Phys. G 49, 055101
(2022).





\bibitem{bc-1-2}
 E. Eichten and F. Feinberg. Spin dependent forces in
qcd. Phys. Rev., D23, 2724, (1981).
W.-K. Kwong, J. L. Rosner. Masses of
new particles containing b quarks. Phys. Rev., D44, 212, (1991).


\bibitem{exotic}
Luciano Maiani
 and Alessandro Pilloni,
GGI Lectures on Exotic Hadrons,	arXiv:2207.05141 [hep-ph]


\bibitem{B-mixing}
J. Vijande, A. Valcarce  and F. Fernandez,
B meson spectroscopy,
Phys. Rev. D77, 017501 (2008)

\bibitem{exotic-pseudosca}
Teng Ji, Xiang-Kun Dong, Feng-Kun Guo, and Bing-Song Zou,
Prediction of a Narrow Exotic Hadronic State with Quantum Numbers
$J^{PC}=0^{--}$,
Phys. Rev. Lett. 129, 102002 (2022)
arXiv:2205.10994v1 [hep-ph]



\bibitem{mass-shifts-channel}
A. M. Badalian, Yu.A. Simonov, and M. A. Trusov,
Chiral transitions in heavy-light mesons,
Phys. Rev. D77, 074017 (2008)





\bibitem{scalars-light}
 J. R. Pelaez,   
From controversy to precision on the sigma meson: a review on the status of
 the non-ordinary $f_0(500)$ resonance, 
 Phys. Rep. 658, 1 (2016).




\bibitem{NJL-scalars}
F.L.Braghin,
Quark-antiquark states of the lightest scalar mesons within the
  Nambu-Jona-Lasinio model with flavor-dependent coupling constants,
 arxiv[hep-ph]
2212.06616.



\bibitem{guo-etal-2008}
Feng-Kun Guo, Siegfried Krewald, Ulf-G. Meissner,
Hadronic-loop induced mass shifts in scalar heavy–light mesons,
Physics Letters B 665, 157 (2008).



\bibitem{PS-mixing}
E.Witten, Current algebra theorems for the U(1) "Goldstone
boson", Nucl. Phys. B156, 269 (1979); G. Veneziano, U(1)
without instantons, Nucl. Phys. B159, 213 (1979).

\bibitem{thooft}
G.
't Hooft, Computation of the quantum effects due to a
four-dimensional pseudoparticle, Phys. Rev. D 14, 3432
(1976); Erratum, Phys. Rev. D 18, 2199 (1978).


\bibitem{bernard-etal-1987}
V. Bernard, R.L. Jaffe and U.-G. Meissner, 
Strangeness mixing and quenching in the Nambu-Jona-Lasinio model,
Nucl. Phys. B
308,  753 (1987).




\bibitem{dimitrasinovic-hl}
V. Dmitrasinovic,
Chiral symmetry of heavy-light scalar mesons with UA(1)
 symmetry breaking,
Phys. Rev. D86, 016006 (2012)


\bibitem{creutz}
M. Creutz, Ann. Phys. (Amsterdam) 323, 2349 (2008).

 

\bibitem{BFM}
L. F. Abbott, 
 Introduction to the background field method,
Acta Phys. Pol. B 13,  33 (1982).


\bibitem{PRD-2014}
A. Paulo Jr., F.L. Braghin, Phys. Rev. {\bf D 90} , 014049  (2014).



\bibitem{PLB-2016}  
F.L. Braghin, 
SU(2) Higher-order effective quark interactions from polarization, 
 Phys. Lett. {\bf B 761}, 424 (2016).







\bibitem{PDG}
R.L. Workman et al. (Particle Data Group),  Prog.Theor.Exp.Phys. 2022, 083C01  (2022).
 K. Nakamura et al. (Particle Data Group),  J. Phys. G 37,
075021 (2010) .
M. Tanabashi et al. (Particle Data Group), 
 Phys. Rev. D 98,   030001  (2018).




\bibitem{EPJA-2016}
 F.L. Braghin, 
Quark and pion effective couplings from polarization effects,
 Eur. Phys. Journ. {\bf A 52},  134  (2016).


\bibitem{PRD-2019}
F.L. Braghin, 
Pion constituent quark coupling strong form factors: a dynamical approach,
Phys. Rev. {\bf D 99},  014001 (2019).



\bibitem{weinberg}
 S. Weinberg, The Quantum Theory of Fields vol. 2: 
Modern Applications, (Cambridge University Press, Cambridge, 1996).



\bibitem{EPJA-2018}
F.L. Braghin,
Low energy constituent quark and pion effective couplings in a weak external
magnetic field,
Eur. Phys. Journ. {\bf A 54}, 45 (2018).
 


\bibitem{mosel}
U. Mosel,  
Path Integrals in Field Theory,
An Introduction,  (2004)
Springer.




\bibitem{YuMod-2021}
A.A. Nogueira, F.L. Braghin, Spontaneous symmetry breakings in the singlet scalar Yukawa model within the auxiliary field method, 
International Journal of Modern Physics A37,  2250066 (2022);
arXiv:2103.05133.

 


\bibitem{sakurai}
J.J. Sakurai  and  S.F. Tuan,  Modern Quantum Mechanics (Addison-Wesley) (1985).

 


\bibitem{klimt-etal}
S. Klimt et al, Generalized SU(3) Nambu-Jona-Lasinio model (I),
Nucl. Phys. A516, 429 (1990).






\bibitem{PNJL-D}
D. Blaschke,  P. Costa, Yu. L. Kalinovsky,
D mesons at finite temperature and density in the 
Polyakov–Nambu–Jona-Lasinio model,
Phys. Rev. D85, 034005 (2012).




\bibitem{cutoff-IR}
D. Ebert, T. Feldmann, H. Reinhardt,
Extended NJL model for light and heavy mesons without q$\bar{q}$
thresholds,
Phys. Lett.
B 388,  154 (1996).



\bibitem{narison-Fps} 
Stephan Narison,
Decay Constants of Heavy-Light Mesons from QCD,
Nuclear and Particle Physics Proceedings, 1 (2015).


\bibitem{narison-Fps2}
Stephan Narison,
A fresh look into $\bar{m}_{c,b}(\bar{m}_{c,b})$
 and precise $f_{D(s),B(s)}$ from heavy–light QCD
spectral sum rules,
Phys. Lett. B 718, 1321 (2013).



\bibitem{HM-decay-lucha-etal}
Wolfgang Lucha , Dmitri Melikhov , and Silvano Simula
Heavy-Meson Decay Constants:
QCD Sum-Rule Glance at Isospin Breaking
EPJ Web of Conferences 129,
QCD-Work 2016
00026 (2016).



\bibitem{mixing-etas}
Th. Feldmann, P. Kroll, B. Stech,
Mixing and Decay Constants of Pseudoscalar Mesons,
arXiv:hep-ph/9802409v2


\bibitem{extrapolation-Fps}
Xin-Heng Guo, Ming-HuaWeng,
Chiral extrapolation of lattice data for B-meson decay constant
Eur. Phys. J. C 50, 63 (2007).



\bibitem{flag}
S. Aoki et al, FLAG Review 2019, EPJ {\bf C 80}, 2 (2020).



\bibitem{lightcondensate}
C. McNeile, {\it et al},
Direct determination of the strange and light quark condensates from full lattice QCD,
arXiv:1211.6577v1 [hep-lat]


\bibitem{nonvalence}
Adam Szczepaniak, Chueng-Ryong Ji, and Stephen R. Cotanch,
Significant Nonvalence Components in Heavy Quark Systems,
Phys. Rev. Lett. 72, 2538  (1994).

\bibitem{charge-radia}
For example in 
C.W. Hwang, Charge radii of light and heavy mesons,
 Eur. Phys. J. C23, 585 (2002).




\bibitem{pelaez-status}
J.R. Pelaez, (2014
Status of light scalar mesons as non-ordinary mesons,
 Journ. of Phys.: Conference Series 562, 012012.

 





 

 

\end{thebibliography}
\end{document}